# Compressive Measurements Generated by Structurally Random Matrices: Asymptotic Normality and Quantization

Raziel Haimi-Cohen and Yenming Mark Lai

*Abstract*— Structurally random matrices (SRMs) are a practical alternative to fully random matrices (FRMs) when generating compressive sensing measurements because of their computational efficiency and their universality with respect to the sparsifing basis. In this work we derive the statistical distribution of compressive measurements generated by various types of SRMs, as a function of the signal properties. We show that under a wide range of conditions, that distribution is a mixture of asymptotically multi-variate normal components. We point out the implications for quantization and coding of the measurements and discuss design consideration for measurements transmission systems. Simulations on real-world video signals confirm the theoretical findings and show that the signal randomization of SRMs yields a dramatic improvement in quantization properties.

## I. INTRODUCTION

Compressive sensing [1] is concerned with determining a signal $\mathbf{x} \in \mathbb{R}^n$ from a vector of measurements,

$$\mathbf{y} = \Phi \mathbf{x} \tag{1}$$

where $\Phi \in \mathbb{R}^{m \times n}$, $m \ll n$, is a *sensing matrix*, and $\mathbf{x}$ is $k$-sparse representation in the column space of a *sparsifyer* $\Psi$,

$$\mathbf{x} = \Psi \zeta, \quad \|\zeta\|_0 \leq k, \tag{2}$$

where $\Psi$ is an orthogonal or a tight frame matrix and $\|\zeta\|_0$ denotes the number of non-zero entries in $\zeta$. If $\Phi\Psi$ meets certain conditions, $\zeta$ and hence $\mathbf{x}$ can be reconstructed from $\mathbf{y}$ by solving the constrained minimization problem

$$\min \|\zeta\|_1 \quad \text{s.t.} \quad \mathbf{y} = \Phi\Psi\zeta \tag{3}$$

Other results in the same vein extend the results to compressible signals (signals which can be approximated by sparse signals), or provide error bounds on the reconstructed solution when the measurements contain noise (In this case (3) may also be modified to account for the noise).

R. Haimi-Cohen is with Alcatel-Lucent Bell-Laboratories, Murray Hill, NJ 07974, USA (phone: 908-582-4159; e-mail: razi@ alcatel-lucent.com).

Y. M. Lai is with University of Texas, Austin, TX 78712 USA (e-mail: mlai@ices.utexas.edu).

## A. Sensing Matrix Design

Various design methods attempt to generate a sensing matrix $\Phi$ that enables correct reconstruction of **x** from a small number of measurements in a computationally efficient way. Generally this goal is achieved only with very high probability (w.h.p.): either $\Phi$ is a random matrix and w.h.p., the selected instance of $\Phi$ enables correct and efficient reconstruction of every possible $(\mathbf{x}, \zeta)$ pair which satisfy (2); or **x** and $\zeta$ are random signals which satisfy (2) and $\Phi$ is deterministic such that the pair $(\mathbf{x}, \zeta)$ can be reconstructed efficiently w.h.p. [2]. In this paper, we are interested in the first option.

A *fully random matrix* (FRM) is a matrix whose entries are independent, identically distributed (IID) Gaussian or Bernoulli random variables (RVs) [3][4]. If $m \geq O(k \log(n/k))$, then for any given $\Psi$, w.h.p., $\Phi\Psi$ is such that every **x** and $\zeta$ which satisfy (2) can be reconstructed by solving (3). FRMs are *universal*, that is, the design of $\Phi$ is independent of $\Psi$, hence the choice of sparsifier can be deferred to the reconstruction stage, which is of significant practical importance. However, because of their completely unstructured nature, FRMs are computationally unwieldy in large scale applications since the random matrix needs to be both computed and stored.

*Randomly sampled transforms* (RST) address the computational complexity problem by imposing structural constraints on the randomness. Let

$$\Phi = \sqrt{n/m} SW$$

where $W \in \mathbb{R}^{n \times n}$ is a square, orthonormal matrix having a fast transform, and $S \in \mathbb{R}^{m \times n}$ is a *random entries selection matrix*, that is, a matrix whose rows are selected randomly, with uniform distribution, from the rows of $I_n$, the $n \times n$ identity matrix. $\Phi\mathbf{x}$ can then be computed efficiently by calculating the fast transform $W\mathbf{x}$ and selecting a random subset of the transform coefficients. RSTs guarantee a correct solution, w.h.p., if

$$m \geq O\left(\mu^2(W, \Psi) k \log n\right) \qquad (4)$$

where $\mu(W, \Psi)$, the mutual coherence of $W$ and $\Psi$, is:

$$\mu(W, \Psi) \triangleq \sqrt{n} \max_{1 \leq i \leq m, 1 \leq j \leq n} |\mathbf{w}_i \mathbf{\psi}_j| / \left(\|\mathbf{w}_i\|_2 \|\mathbf{\psi}_j\|_2\right)$$

where $\mathbf{w}_i$, $\mathbf{\psi}_j$ are the *i*th row and *j*th column of $W$, $\Psi$, respectively [5]. Since $1 \leq \mu(W, \Psi) \leq \sqrt{n}$, we can choose $m \ll n$ which satisfies (4) only if $W$ is selected so that $\mu(W, \Psi)$ is small. Therefore, RSTs are not universal.

The universality issue was addressed by the introduction of *structurally random matrices* (SRM) [6][7]:

$$\Phi = \sqrt{n/m} SWR \qquad (5)$$

where $S, W$ are as above and $R \in \mathbb{R}^{n \times n}$, the *randomizer*, is a random matrix. Hence

$$\Phi \mathbf{x} = \sqrt{n/m} SW(R\mathbf{x}) = \sqrt{n/m} SW(R\Psi)\zeta .$$

Therefore, a SRM with a given sparsifier $\Psi$ behaves as the RST $\sqrt{n/m}SW$ with the random sparsifier $R\Psi$. If $R\Psi$ and $W$ are mutually incoherent w.h.p., then SRMs are universal, and the known results for RSTs with incoherent sparsifiers (e.g. performance with compressible signals or noisy measurements) hold w.h.p.

Two types of randomization were proposed: *Local randomization* (LR), where each entry of **x** is multiplied by ±1 with equal probability; and *global randomization* (GR), where the entries of **x** are randomly shuffled. Both forms are computationally simple and were shown, for a large class of transforms $W$, to be universal [7]. The universality of LR was extended in [9] to the more general case where $\sqrt{n/m}SW$ in (5) is replaced by any matrix with the restricted isometry property (RIP).

Other methods were also proposed for constructing universal and computationally efficient sensing matrices, such as Random convolution (RC) [8].

*B. Quantization and coding of measurements*

Many application of compressive sensing, e.g. video surveillance and streaming [10]–[16] involve sending the measurement for processing over a communication channel. The transmission of measurements requires a *coding scheme*, which entails source coding that is typically implemented by quantization followed by channel coding of the quantization codewords.

Conventional media coding standards are efficient over a wide range of input signals and operating conditions. One of the keys to this robustness is the usage of various signal-adaptive techniques in order to control the bit rate and improve performance. These techniques are applied before, during, and after quantization. For example, a linear prediction[17][18] model may be estimated for the signal and the quantization may be performed on the prediction error, which reduces the bit rates needed to achieve specific quantization accuracies; the granularity of the quantizer may be varied according the signal content; and one out of several possible variable length coding schemes may be selected to achieve low rate lossless coding of the quantization codewords. The parameters of the linear prediction model, the quantizer granularity, and the lossless coding scheme need to be shared with the decoder, and hence they are encoded and sent as side information. Since the side information is critical for the decoding of the signal as a whole, it is typically encoded with higher accuracy and, in noisy channels, with better error protection, than the rest of the data. The amount of data in the side information is very small, hence the bit rate overhead caused by sending it is usually negligible in comparison to the performance achieved by it.

The preferred coding scheme for compressive measurements depends on a variety of factors, but it is invariably based on assumptions about the probability distribution of the measurements, which is determined by the type of sensing matrix used. Furthermore, applying any of the signal-adaptive techniques described above requires having a parametric model where this distribution is characterized by parameters estimated from the signal and transmitted to the decoder as side information.

The quantization of compressive measurements has recently received significant attention. Dai et al. [19][20] studied the effect of quantization on reconstruction accuracy with various quantizer designs and provided asymptotic boundaries on the rate-distortion function when quantization is followed by Huffman coding [20]. The efficacy of uniform vs. non-uniform scalar measurement quantization was compared specifically for video signals in [14][21]. Unlike all other quantizer designs we reviewed, the quantizer of [14] is signal-adaptive: its operation is controlled by the variance of the measurements in each frame, which is sent to the decoder as side information. A quantizer optimized for compressed sensing reconstruction is presented in [22]. Laska et al. studied the effect of saturation [23], the trade-off between the number of measurements and quantization accuracy [24] and the extreme case of 1-bit quantizers [24]–[26]. Modifications to the reconstruction algorithms to address quantization effects were proposed in [23][26]. In all these papers, the sensing matrix was a FRM, and in many of them the input signal was assumed to be random with a known distribution. Also, in most cases there was no attempt to reduce the bit rate of the quantizer codewords by applying variable rate coding. Therefore, those results cannot be applied to the design of signal adaptive coding schemes for measurements generated by SRMs.

Do et al. [7] showed that under certain conditions, the entries of $\sqrt{n/m}WR\Psi$ are asymptotically univariate normal, where $\Psi$ is an orthonorma matrix. With proper scaling, **x** may be viewed as a column of $\Psi$, which gives us conditions for asymptotic univariate normality of the entries of $\sqrt{n/m}WR\mathbf{x}$. We did not find similar results for RC. In addition, upper bounds on the tail distribution probability of these entries have been derived (albeit "buried" as intermediate steps in proofs of lemmas) for both SRM in [7] and RC [8]. However, [7] and [8] do not provide explicit expressions for the variance of those entries, hence these results are of limited use for actual quantizer design.

*C. Our contribution*

In this paper we study the distribution of measurements generated by SRMs and draw conclusions for the design of coding schemes for SRM generated measurements. We demonstrate our results by simulation on real-world signals and show that SRMs are superior to RSTs not only in being universal, but also by the fact that the measurements' distributions lend themselves to effective quantization. In addition, while [7] states that SRM is distinct from RC by virtue of the signal randomization step, we show that in fact, RC may be viewed as frequency domain variant of LR-SRM (with randomization in the frequency domain), with similar measurements distribution properties.

The design of a coding scheme for compressive measurements depends on application specific considerations. Our goal here is not to propose a particular coding scheme, but to study the properties of the measurements' distribution and to point out ways in which these properties may be utilized in designing such a coding scheme. Each measurement is a mixture of the random entries of $\sqrt{n/m}WR\mathbf{x}$. These entries may be highly correlated and not identically distributed. Our results enable the designer to optimize the coding scheme by taking into account the individual distribution of each of these entries as well as the correlation among them. In addition, we show how the signal-adaptive information may be represented by parameteric models with a small number of parameters that can be shared with the decoder as side information. In particular, we show that in local randomization, the entries of $\sqrt{n/m}WR\mathbf{x}$ are generally highly correlated, whereas in many important cases of global randomization, those entries are uncorrelatd and asymptotically independent.

Our work expands the univariate asymptotic normality results in [7] in the following ways: We extend those results to the multivariate case; we provide similar multivariate asymptotic normality results for RC; and, in the univariate case, we relax the conditions for asymptotic normality. Our results on bounds for the tail distribution provide a stronger bound than that found in [8] for the RC case, and in the GR case, remove some restrictive conditions on $W$ and $\mathbf{x}$ in which were required by [7]. In addition, we provide explicit expressions for the means and covariances of the measurements and we present results about asymptotic independence and uniform convergence to normality, which are necessary to justify the approximations for large $n$ in the coding schemes.

This work differs from previous work on compressive measurement quantization in the following aspects: We study measurements generated by practical sensing matrices, such as SRMs and the closely related RC; we provide a rigorous characterization of the measurements distribution, including cross-correlation between measurements; and we propose methods for signal-adaptive quantization and channel-coding of measurements. For the latter, we assess the amount of side information needed by the decoder.

The paper is organized as follows: Sec. II introduces notations and concepts which are common to both LR and GR SRMs and provides examples of transforms suitable for use with SRMs. Sec. III introduces methods for SRMs measurement quantization and channel coding. Sec. IV and Sec. V characterize the distribution of measurements generated by SRMs with LR and GR, respectively. Sec. VI shows simulated SRMs measurements distributions and Sec. VII presents conclusions.

## II. GENERAL PROPERTIES AND EXAMPLES OF SRMS

### A. Notation

Let $\Phi^{(n)} \triangleq \sqrt{n/m}S^{(n)}W^{(n)}R^{(n)} \in \mathbb{R}^{m \times n}$ be a SRM as defined by (5) and denote the $j$th row of $W^{(n)}$ by $\mathbf{w}_j^{(n)} \triangleq [w_{j1}^{(n)},\ldots,w_{jn}^{(n)}]$. We assume that $R^{(n)} \in \mathbb{R}^{n \times n}$ is a random orthonormal matrix. Let $\{x_k\}_{k=1,2,\ldots}$ be a deterministic bounded signal, $|x_k| \leq x_{\max}$ and let $\mathbf{y}^{(n)} = \Phi^{(n)}\mathbf{x}^{(n)}$ be a vector of $m$ compressive sensing measurements obtained from the $n$-dimensional signal vector $\mathbf{x}^{(n)} \triangleq [x_1,\ldots,x_n]^T$. Often the fast transform $W^{(n)}$ exists only for specific orders (e.g. powers of 2). If the source signal is finite and of a dimension for which the transform does not exist, we assume that the signal is zero padded to the next available transform order.

Let $\mathbf{z}^{(n)} \triangleq \sqrt{n/m}W^{(n)}R^{(n)}\mathbf{x}^{(n)}$ be a random variable (RV) defined on the probability space of the randomizer $R^{(n)}$. For $1 \leq j, h \leq n$, define:

$$\mu_{jn} \triangleq E\{z_j^{(n)}\}$$

$$\text{cov}\{z_j^{(n)}, z_h^{(n)}\} \triangleq E\{z_j^{(n)} z_h^{(n)}\} - \mu_{jn}\mu_{hn} \tag{6}$$

$$\sigma_{jn}^2 \triangleq \text{var}\{z_j^{(n)}\} = \text{cov}\{z_j^{(n)}, z_j^{(n)}\} \tag{7}$$

If $\mathbf{v}$ is a $r$-dimensional random vector, $\text{cov}\{\mathbf{v}\}$ is the $r \times r$ matrix defined by

$$\text{cov}\{\mathbf{v}\}_{i,j} \triangleq E\{v_i v_j\} - E\{v_i\}E\{v_j\}, \quad 1 \leq i, j \leq r.$$

By substitution, $\mathbf{y}^{(n)} = S^{(n)}\mathbf{z}^{(n)}$ and since $S^{(n)}$ is a random entry selection matrix, the compressive sensing measurements are RVs given by $y_k^{(n)} = z_{c_n(k)}^{(n)}$, $1 \leq k \leq m$, where the *measurements indices* $c_n(1), \ldots, c_n(m)$ are a random sample from $\{1, \ldots, n\}$, *with or without replacement*, that is, $\mathbf{c}_n \triangleq [c_n(1), \ldots, c_n(m)]^T$ is uniformly distributed, either in $C_0^{(n)} \triangleq \{1, \ldots, n\}^m$, for sampling with replacement or in $C_1^{(n)} \triangleq \{\mathbf{g} = [g_1, \ldots, g_m]^T \in C_0 \mid g_1, \ldots, g_m \text{ are distinct}\}$, for sampling without replacement. Each of the measurements $y_1^{(n)}, \ldots, y_m^{(n)}$ is a mixture, with equal probabilities, of the *mixture components* $z_1^{(n)}, \ldots, z_n^{(n)}$, hence $y_1^{(n)}, \ldots, y_m^{(n)}$ are identically distributed. When discussing the measurements' distribution, $y^{(n)}$ stands for any one of $y_1^{(n)}, \ldots, y_m^{(n)}$. Thus,

$$\mu_{yn} \triangleq E\{y^{(n)}\} = n^{-1}\sum_{j=1}^{n} z_j^{(n)} \tag{8}$$

$$\sigma_{yn} \triangleq \text{var}\{y^{(n)}\} = E\{(y^{(n)})^2\} - \mu_{yn}^2 \tag{9}$$

$\langle a \rangle_n$ denotes the value $0 < a' \leq n$ such that $a \equiv a' \mod(n)$. $\mathbf{1}_n, \mathbf{0}_n \in \mathbb{R}^n$ are the vectors $[1, \ldots, 1]^T$, $[0, \ldots, 0]^T$ respectively.

When discussing asymptotic behavior as $n \to \infty$ the notation is usually simplified, without loss of generality, by not indicating that $n$ is restricted to orders at which the transform exists and that $m$ is a function of $n$. If $n$ is fixed we usually omit the superscript $(n)$ and subscript $n$ for clarity.

For vectors $\mathbf{u}, \mathbf{v} \in \mathbb{R}^n$, $\mathbf{v} * \mathbf{u}$ and $\mathbf{v} \circ \mathbf{u}$ denote convolution and pointwise multiplication, respectively, that is

$$(\mathbf{v} * \mathbf{u})_j \triangleq \sum_{k=1}^{n} v_k u_{\langle j-k \rangle_n}, \quad 1 \leq j \leq n$$

$$(\mathbf{v} \circ \mathbf{u})_j \triangleq v_j u_j, \quad 1 \leq j \leq n$$

Let $r > 0$ be a fixed integer. A sequence $\{\mathbf{u}^{(n)}\}$ of $r$-dimensional RVs is *asymptotically multivariate normal* (AMN) if for sufficiently large $n$, $\text{cov}\{\mathbf{u}^{(n)}\}$ exists, is positive definite, and

$$\left(\text{cov}\{\mathbf{u}^{\{n\}}\}\right)^{-1/2}\left[\mathbf{u}^{(n)} - E\{\mathbf{u}^{(n)}\}\right] \xrightarrow[n\to\infty]{d} \mathcal{N}(\mathbf{0}_r, I_r).$$

*B. Statistical Properties*

If $c_n(1),\ldots,c_n(m)$ are selected with replacement, then the measurements are independent. However, in many practical applications, they are selected without replacement, i.e. $c_n(1),\ldots,c_n(m)$ are distinct. In that case, for large $n$, if $m \ll n$ the measurements are approximately independent, as stated by the Lemma 1 below:

*Lemma 1*: Let $m = m(n) = o(\sqrt{n})$ and let $F_0^{(n)}, F_1^{(n)}$ be the multivariate distribution functions of $\mathbf{y}^{(n)}$ with and without replacement, respectively. Then for any sequence $\mathbf{h}^{(n)} \in \mathbb{R}^{m(n)}$, $n=1,\ldots$, such that $\liminf_{n\to\infty} F_0^{(n)}(\mathbf{h}^{(n)}) > 0$,

$$F_1^{(n)}\left(\mathbf{h}^{(n)}\right) \big/ F_0^{(n)}\left(\mathbf{h}^{(n)}\right) \xrightarrow[n\to\infty]{} 1 \quad \square.$$

We characterize the distribution of the measurements in the following steps: First we derive expressions for the expectation and covariances of the mixture components $z_1,\ldots,z_n$. The expressions for the variance of each of the mixture components, $\sigma_1^2,\ldots,\sigma_n^2$, and the mean of each of the compressive measurements, $\mu_y$, follow immediately from (7) and (8), respectively. Since $R$ and $W$ are orthonormal:

$$E\{y^2\} = \frac{1}{n}\sum_{i=1}^{n} Ez_i^2 = \frac{1}{n} E\|\mathbf{z}\|_2^2 = \frac{1}{m} E\{\|WR\mathbf{x}\|_2^2\} = m^{-1}\|\mathbf{x}\|_2^2.$$

Therefore:

$$\sigma_y^2 = m^{-1}\|\mathbf{x}\|_2^2 - \mu_y^2. \tag{10}$$

Then we provide an upper bound on the tail distribution of the mixture components $z_1,\ldots,z_n$.

$$\Pr\{|z_j - \mu_j| > t\sigma_j\} \leq 2\exp(-t^2\eta(t,\tau_j)), \quad t \geq 0 \tag{11}$$

where $\eta(t,\tau_j)$ is continuous function which is slowly decreasing in $t$. $\eta$ and $\tau_j$ are determined by the specific type of randomization and in several important cases $\eta(t,\tau_j)$ is constant or nearly constant in the range of interest for $t$.

Finally, we give conditions for asymptotic normality of the mixture components. For any fixed integer $r > 0$, let $\{\mathbf{u}^{(n)}\}$ be a sequence of $r$-dimensional RVs such that,

For sufficiently large $n$:  $u_k^{(n)} = z_{j_n(k)}^{(n)}$, $1 \leq k \leq r$, and $1 \leq j_n(1),\ldots,j_n(r) \leq n$ are distinct. $\tag{12}$

We show that $\{\mathbf{u}^{(n)}\}$ is AMN if the following two conditions are met:

$$\lim_{n \to \infty} \max_{1 \leq k \leq r} \left\| \mathbf{w}_{j_n(k)}^{(n)} \right\|_\infty = 0 \tag{13}$$

$$\forall \boldsymbol{\alpha} \in \mathbb{R}^r, \boldsymbol{\alpha} \neq \mathbf{0} : \liminf_{n \to \infty} (m/n) \boldsymbol{\alpha}^T \operatorname{cov}\left(\mathbf{u}^{(n)}\right) \boldsymbol{\alpha} > 0 \tag{14}$$

which, for the univaritate case $r = 1$, becomes (for clarity, the index $j_n(1)$ is written as $j_n$):

$$\liminf_{n \to \infty} (m/n) \sigma_{j_n,n}^2 > 0. \tag{15}$$

The univariate asymptotic normality results in [7] (with proper scaling) require the following conditions for the LR case, with some additional assumptions in the GR case:

$$C_1 \leq n^{1/2} \left| \mathbf{w}_{j_n}^{(n)} \right| \leq C_2, \, n = 1, \ldots \tag{16}$$

$$\left\| \mathbf{x}^{(n)} \right\|_\infty / \left\| \mathbf{x}^{(n)} \right\|_2 \to 0 \tag{17}$$

where $C_1, C_2$ are poistive constants. Clearly, the (16) is more restrictive for the transform matrix than (13).

The following lemma guarantees that if there is convergence to normality it is uniform over all choices of $\{\mathbf{u}^{(n)}\}$ which satisfy (12)–(14). Therefore, for sufficiently large $n$, any choice of $\{\mathbf{u}^{(n)}\}$ is arbitrarily close to multivariate normal.

*Lemma 2:* Let $r > 0$ be a fixed integer and let $J^{(n)} \subseteq \{1, \ldots, n\}$, $n = 1, \ldots$ be sets of indices. Suppose that any sequence $\{\mathbf{u}^{(n)}\}$ which satisfies (12) such that $\{j_n(1), \ldots, j_n(r)\} \subseteq J^{(n)}$ is AMN. Then the probability distribution of $\{\mathbf{u}^{(n)}\}$ converges to $\mathcal{N}(\mathbf{0}_r, I_r)$ uniformly over all sets of distinct indices $\{j_n(1), \ldots, j_n(r)\} \subseteq J^{(n)}$ □.

*Examples of Transforms Used by SRMs*

Common examples for the transform matrix $W$ used in (5) are the Discrete Cosine Transform (DCT) and the Walsh-Hadamard Transform (WHT) matrices, each with rows $\mathbf{w}_j$, $j = 1, \ldots, n$ scaled so that $\|\mathbf{w}_j\|_2^2 = 1$ (in our discussion of specific transforms, we always assume this normalization). The Discrete Fourier Transform (DFT) may also be used as long as it is treated as a real (orthonormal) matrix, that is, the real and imaginary parts of each row of the original DFT are separated and normalized to form two unit-norm rows of real coefficients. Then, redundant rows are deleted to make $W$ a $n \times n$ matrix. Transform matrices for multi-dimensional signals may be formed as a Kronecker product of any orthonormal matrices [28].

By construction, WHT, DCT and DFT satisfy condition (13), that limits the concentration of energy within each row of $W$. Also, $\{W^{(n)}\} = \{W'_n \otimes W''_n\}$ satisfies this condition if either $\{W'_n\}$ or $\{W''_n\}$ satisfies it. Condition (13) is a weaker form of the conditions for a SRM with large $n$ to be incoherent with any sparsifier w.h.p. [7]. Therefore transform matrices $W^{(n)}$ for which condition (13) does not hold, such as wavelet transforms or the trivial $W^{(n)} = I_n$, would not make good SRMs to begin with.

III. MEASUREMENT QUANTIZATION AND CHANNEL CODING

*A. Determining Variables for Quantization and Coding*

Since the compressive measurements $y_1,\ldots,y_m$ are, at least approximately, mutually independent, the advantage of vector quantization over scalar quantization is small compared to the cost of added complexity ([20] reaches a similar conclusion for FRM generated measurements). Therefore, we can apply a coding scheme in which each measurement is quantized by the same scalar quantizer and then lossless channel coding is applied to the quantization codewords to represent them as a bit sequence. The parameters of this coding scheme depend on the signal and need to be shared with the decoder. As we will show, under a wide range of conditions the mixture components $z_1,\ldots z_n$ are AMN, and the distribution of each of them is fully characterized by their mean and variance. In some important cases, $z_1,\ldots z_n$ are identically distributed and hence the coding scheme is parametrized by a single mean and variance. If this is not the case, the mixture of $z_1,\ldots z_n$ can be approximated by a mixture of a small number of Gaussians. The means, variances, and weights of the approximate mixture are sent as side information and used by both encoder and decoder to compute the parameters of the coding scheme.

While this method is simple, it is suboptimal if the distributions of $z_1,\ldots z_n$ are different. In this case, a lower bit rate can be achieved at the same distortion level by adapting the coding scheme for each of the measurements $y_k = z_{c(k)}, 1 \leq k \leq m$, to the specific distribution of $z_{c(k)}$. In entropy terms, $H\{y_k\} - H\{y_k | c(k)\} = I\{y_k, c(k)\} >= 0$, where $H, I$ denote entropy and mutual entropy, respectively. The number of bits which can be saved by a coding scheme adapted to the specific $z_{c(k)}$ is $I\{y_k, c(k)\}$ which is zero only if $z_1,\ldots,z_n$ are identically distributed. This requires sharing the coding scheme of each measurement with the decoder. As we will show, under a wide range of conditions the means and variances of $z_1,\ldots z_n$ may be represented by a parametric model. The model parameters may be sent as side information, and both encoder and decoder may derive the coding scheme for each measurement from the mean and variance computed by the model.

Quantizing mixture components, rather than measurements, eliminates the need for independence of the measurements but raises the question of independence among the mixture components. Fortunately, when the latter are AMN, independence can be achieved by decorrelation. One relatively simple way to achieve this is by linear prediction, as outlined in Appendix A. In this approach some mixture components are linearly predicted from other mixture components and the coding scheme is applied to the linear prediction residuals rather than to the original mixture components. If the measurements are highly correlated their variances are much larger than those of the residuals, hence the latter can be coded more efficiently.

In Sec. IV and V below we show how the parameters of the distribution of the mixture components $z_1,\ldots z_n$ can be derived from the properties of the signal. In some applications the measurements are computed in the analog domain and the signal is not available. In such cases, the theoretical results inform us of the appropriate model for the measurements' distribution, and the model parameters may be estimated directly from the available measurements.

*B. Scalar Quantization and Channel Coding*

We turn our attention to the design of a coding scheme based on scalar quantization. The data is a sequence of $m$ *transmitted variables* (TVs), which may be measurements, selected mixture components, or linear prediction residuals. The distributions of the TVs are given by a parametric model and they may or may not be identical. The coding schemes for each TV is adapted to its specific distribution and accordingly, they too may or may not be identical.

A scalar quantizer $Q(y) \triangleq \arg\min_{c \in C} |y - c|$ maps a TV $y$ into a finite *codebook*. The *quantization region* of a *codeword* $c \in C$ is $Q^{-1}(c) \triangleq \{u \in \mathbb{R} \mid Q(u) = c\}$. A codeword $c \in C$ approximates the values in its *quantization region*, $Q^{-1}(c) \triangleq \{u \in \mathbb{R} \mid Q(u) = c\}$ and the codeword distortion is $E\{|y - c|^2 \mid y \in Q(c)\}$. If $Q^{-1}(c)$ is a finite interval $c$ is *unsaturated* and the quantization error in $Q^{-1}(c)$ is bounded; otherwise $c$ is saturated and the quantization error in $Q^{-1}(c)$ is unbounded. The range of the quantizer is the union of all bounded quantization regions. The accuracy of signal reconstruction from quantized measurements is severely degraded if even a small number of input codewords are saturated (the same would be true for measurements derived from saturated TVs, such as residuals). The reconstruction algorithm may be modified to prevent this degradation by a special handling of saturated measurements, but the results are only slightly better than those obtained when saturated measurements are simply discarded [23]. Therefore, the quantizer's range should be wide enough to make saturation a rare event. On the other hand an excessively large quantizer's range yields little performance gain and can add unnecessary complexity. Suppose that an upper bound for saturation frequency is specified as $1 > \delta > 0$. If the same quantizer is used for all TVs, the quantizer range can be set empirically so that no more than $m\delta$ measurements are saturated. However, if different quantizers are used, e.g. if the TVs are mixture components with different variances or linear prediction residuals, then in order to set an appropriate range for each quantizer, one may need to invoke the theoretical results provided here about the measurements' distribution, such as asymptotic normality or the concentration result (11). The latter may be necessary because convergence to normal distribution may be slower at the tail. In addition, those results may be useful in order to assess the ranges that will be needed for a given class of input signals.

After quantization, channel coding is applied to the codewords to represent them as a bit sequence. The entropy $H(Q(y)) \triangleq E\{\log_2 \Pr(Q(y))\}$ is a lower bound on the coding rate, (in bit/measurement units). Methods of variable length coding (VLC), such as arithmetic coding [29] can get arbitrarily close to the entropy rate, provided that the codewords distribution is known. A low complexity alternative to VLC is fixed length coding (FLC), the rate of which is bounded from below by $\log_2 |C|$ bit/measurement. This bound is attainable if $\log_2 |C|$ is an integer. Otherwise, arbitrarily close data rates can be achieved by jointly coding groups of several codewords.

We consider two types of scalar quantizers. First, we consider an optimal quantizer [30] which, based on knowledge of the measurements' distribution, minimizes the distortion subject to a constraint on either the codebook size or the quantizer's entropy. The optimization usually causes the codewords to have similar probabilities, which makes the bit rate of FLC close to that of VLC. Second, we consider a uniform quantizer where the quantization regions of all unsaturated codewords are of the same length. For a memoryless source, under a wide range of conditions, the entropy rate of a uniform quantizer is within a fraction of a bit from that of an optimal quantizer with the same distortion [30][31]. Therefore, there is a trade-off between complex optimal quantization followed by simple FLC versus simple uniform quantization followed by complex VLC.

The measurements may contain noise, e.g. noise propagated from a noisy source signal. With optimal quantization the codeword distortion differs from codeword to codeword. At high bit rates, some codeword distortions may be lower than the measurements' noise floor, effectively wasting bits on representing the noise, while at low bit rates, the distortions of some unsaturated codewords may be large enough to induce severe reconstruction degradation as seen with saturated codewords. Therefore, uniform quantization with VLC appears to be a more robust approach.

Since the coding scheme is signal-adaptive, its settings must be shared with the decoder as side information. Whether we use an optimal quantizer with FLC or a uniform quantizer with VLC, this may require transmitting $|C|$ parameters to specify the values, or the probabilities, of each codeword, respectively. That may be acceptable if the codebook is small, but for a large codebook this amount of side information may be too much. Instead, the measurements' distribution may be modeled by a parametric model, with the side information consisting only of estimates of the model parameters. Both encoder and decoder can then derive the coding scheme from the parametric distributions. Some of the estimated parameters, e.g. the signal mean, may, in fact, be compressive measurements and may be used as such. The designation of values as side information indicates that those values are used to specify the coding scheme and therefore are treated differently: they are guaranteed to be transmitted regardless of the random measurements selection, and they are often transmitted with higher accuracy or more error protection than the rest of the measurements.

## IV. LOCAL RANDOMIZATION

A locally randomized SRM (LR-SRM) is a SRM as defined in (5) whose randomizer is the diagonal matrix

$$R = \begin{pmatrix} b_1 & 0 & \ldots & 0 \\ 0 & b_2 & & 0 \\ \vdots & & \ddots & \vdots \\ 0 & \ldots & 0 & b_n \end{pmatrix}$$

where $b_1,...,b_n$ are IID Rademacher RVs ($b_k = \pm 1$ with equal probability).

## A. Measurements' Distribution with LR-SRM

*Theorem 1 (distribution of LR-SRM mixture components)* [1]: With local randomization, for $1 \leq j, h \leq n$:

(a) The distribution of each mixture components $z_j$ is symmetric and

$$E\{z_j\} = 0, \qquad (18)$$

$$\text{cov}\{z_j, z_h\} = \frac{n}{m}\sum_{k=1}^{n} w_{jk} w_{hk} x_k^2 = \frac{n}{m}(\mathbf{w}_j \circ \mathbf{w}_h)(\mathbf{x} \circ \mathbf{x}), \qquad (19)$$

$$\sigma_j^2 = \text{cov}\{z_j, z_j\} = \frac{n}{m}\sum_{k=1}^{n} w_{jk}^2 x_k^2. \qquad (20)$$

(b)

$$\Pr\{|z_j| > t\sigma_j\} \leq 2e^{-t^2/2}. \qquad (21)$$

(c) For any fixed integer $r > 0$ a sequence $\{\mathbf{u}^{(n)}\}$ defined by (12) is AMN if conditions (13) and (14) are satisfied □.

By (18) and (10), $\mu_y = 0$ and $\sigma_y^2 = m^{-1}\|\mathbf{x}\|_2^2$. Equation (19) shows that if $\mathbf{x}$ is constant then $z_j, 1 \leq j \leq n$ are uncorrelated, but otherwise they can be highly correlated. Since the rows of $W$ form a basis, $\mathbf{w}_j \circ \mathbf{w}_h$ can be represented by a linear combination of $\mathbf{w}_1,\ldots,\mathbf{w}_n$. Often the number of non-zero terms in this represetnation is very small hence (19) can be simplified by substituting

$$\mathbf{w}_j \circ \mathbf{w}_h = n^{-1/2}\sum_{k=1}^{p} \gamma_k(j,h)\mathbf{w}_{l_k(j,h)} \qquad (22)$$

where $p$ is the number of non-zero terms and $\gamma_k(j,h)$, $l_k(j,h)$, $k=1\ldots p$, are weights and indices, respectively. For the WHT, $p=1$, $\gamma_1(j,h)=1$ and $l_1(j,h)$ is derived by bitwise modulo-2 addition of the binary representations of $j$ and $h$. For the DCT and DFT, $p=2$, $\gamma_k(j,h) = \pm 2^{-1/2}$, and $\mathbf{w}_{l_k(j,h)}$ correspond to frequencies which are sums or differences of the frequencies of $\mathbf{w}_j$ and $\mathbf{w}_h$. Therefore, by (19) the covariance can be computed from $W(\mathbf{x} \circ \mathbf{x})$, the transform of $\mathbf{x} \circ \mathbf{x}$, as:

$$\text{cov}(z_j, z_h) = (n^{1/2}/m)\sum_{k=1}^{p} \gamma_k(j,h)\mathbf{w}_{l_k(j,h)}(\mathbf{x} \circ \mathbf{x}) \qquad (23)$$

Furthermore, it can also be shown that if (22), (23) hold for transforms $W', W''$ with $p = p', p''$ in the right hand side, respectively, then (22), (23) hold for the rows of $W' \otimes W''$ with $p = p'p''$.

---

[1] Part (b) may be derived from an intermediate step in the proof of Theorem III.3 in [7]. The univariate case ($r=1$) of part (c) appears as Theorem III.1 in [7], with (13) and (15) replaced by (16) and (17). It is easy to verify that (16) and (17) are narrower and imply (13) and (15).

If $\mathbf{x}$ is a typical media signal, $W(\mathbf{x} \circ \mathbf{x})$ is often highly compressible and hence can be approximated by a model with a small number of parameters. For example, the approximation can be found by first saving a few dominant entries of $W(\mathbf{x} \circ \mathbf{x})$ and setting the rest to zero. For the WHT, DFT and DCT, $\{l_k(j,1),\ldots,l_k(j,n)\} = \{1,\ldots,n\}$ for $1 \le j \le n$, $1 \le k \le p$. Therefore in the WHT case, by (23), each saved entry of $W(\mathbf{x} \circ \mathbf{x})$ determines one entry in each of the $n$ rows of $\text{cov}(z_j, z_h)$. For the DCT and DFT, for each $j$, each saved entry of $W(\mathbf{x} \circ \mathbf{x})$ determines one of the two summands in (23) for two entries in each row of $\text{cov}(z_j, z_h)$. Therefore, the signal covariance may be estimated by (23), using an approximation of $W(\mathbf{x} \circ \mathbf{x})$.

Condition (14) can be written as:

$$\forall \boldsymbol{\alpha} \in \mathbb{R}^r, \boldsymbol{\alpha} \ne \mathbf{0}: \quad \liminf_{n \to \infty} \left\| \left( \sum_{k=1}^r \alpha_k \mathbf{w}_{j(n,k)}^{(n)} \right)^T \circ \mathbf{x} \right\|_2^2 > 0$$

which, roughly speaking, requires that any linear combination of the relevant rows should have some of its energy in areas where the signal has energy.

If all the entries of $W$ have the same magnitude, as with the WHT, The $z_1, \ldots z_n$ have identical univariate distributions with $\sigma_j^2 = m^{-1} \|\mathbf{x}\|_2^2$, $1 \le j \le n$, (13) is satisfied by design, and for $r=1$, (15) becomes a condition on the signal norm:

$$\liminf_{n \to \infty} n^{-1} \|\mathbf{x}^{(n)}\|_2^2 > 0. \tag{24}$$

Suppose that the entries of $W$ do not have the same magnitude, and for some $n$, $z_1, \ldots, z_n$ are approximately univariate normal. Then the tail distribution is determined by the largest among the mixture components variances $\sigma_1^2, \ldots, \sigma_n^2$. This is bounded because, by (19),

$$\sigma_j^2 \le (n/m) \|\mathbf{w}_j\|_\infty^2 \|\mathbf{x}\|_2^2 = \left(n \|\mathbf{w}_j\|_\infty^2\right) \sigma_y^2,$$

In both the DFT and DCT cases, $\|\mathbf{w}_j\|_\infty \le \sqrt{2/n}$, hence $\sigma_j^2 \le 2\sigma_y^2$.

Furthermore, in the DFT case, let $(\mathbf{w}_r, \mathbf{w}_i)$ be a pair of rows which correspond to the real and imaginary parts, respectively, of the $h$th complex DFT coefficient, with DFT coefficient numbering starting at 1. Calculation using (19) yields

$$\sigma_r^2 = \sigma_y^2 + q_h, \quad \sigma_i^2 = \sigma_y^2 - q_h \tag{25}$$

where

$$q_h \triangleq \frac{1}{m} \sum_{k=1}^n x_k^2 \cos \frac{4\pi(h-1)(k-1)}{n}$$

Let $g$ be the greatest common divisor of $n$ and $2(h-1)$ and let $d \triangleq n/g$. $q_h = \sigma_y^2$ or $q_h = -\sigma_y^2$ only if $x_k = 0$ when $\langle k \rangle_d \ne 1$ or $\langle k \rangle_d \neq (1+d/2)$, respectively. In both cases, $\{x_k\}$ is a train of spikes at intervals of $d$ with a specific phase. In order for $q_h$ or $-q_h$ to be close to $\sigma_y^2$, the signal has to approximate such a spike train, that is, $x_k$ must be near zero except when $(\langle k \rangle_d - 1)/d \approx 0$ or when $(\langle k \rangle_d - 1)/d \approx 0.5$, respectively. While such signals are certainly possible, in many practical applications they are uncommon, in which cases $|q_h| \ll \sigma_y^2$ and therefore, $\sigma_r^2, \sigma_i^2 \approx \sigma_y^2$.

*B. Measurements' Distribution with Random Convolution*

Let $F^{(n)}$ be the $n$th-order complex DFT matrix, given by $f_{kj}^{(n)} \triangleq n^{-1/2} \exp[-2\pi i (k-1)(j-1)/n]$. In the following, the superscript $^{(n)}$ is generally omitted for clarity. Note that indexing starts at 1 and $F^{-1} = F^*$, which makes the notation a little different from the usual DFT notation. Let $R$ be a random diagonal matrix with diagonal elements $b_1, \ldots, b_n$ such that $F^* R F$ is real. The RC sensing matrix is [8]:

$$\Phi = \sqrt{n/m} S F^* R F$$

where $S \in \mathbb{R}^{m \times n}$ is a random selection matrix as in (5). The name "random convolution" is because, by the properties of DFT,

$$\mathbf{z} \triangleq \sqrt{n/m} F^* R F \mathbf{x} = \sqrt{n/m} F^* (\mathbf{b} \circ (F\mathbf{x})) = m^{-1/2} (F^* \mathbf{b}) * \mathbf{x} \tag{26}$$

where $\mathbf{b} \triangleq [b_1, \ldots, b_n]^T$. However, since

$$\mathbf{y} = \Phi \mathbf{x} = \sqrt{n/m} S F^* R (F\mathbf{x}) \tag{27}$$

we can think of RC as a variant of LR-SRM, where $W = F^*$ and the signal is $F\mathbf{x}$ — the DFT representation of the original signal. The RVs $b_1, \ldots, b_n$ are uniformly distributed and have unit magnitude, but since $F^* R F$ is required to be real, the sequence $b_1, \ldots, b_n$ must be conjugate symmetric: $b_k = \overline{b}_{\langle n+2-k \rangle_n}$, $k = 1, \ldots, n$. Therefore $b_1, \ldots, b_n$ cannot be IID Rademacher. Let $\chi_n(k)$ be the indicator function for self-symmetric terms in $b_1, \ldots, b_n$, that is, $\chi_n(k) \triangleq 1$ if $k = \langle n+2-k \rangle_n$ (this is true for $k=1$ and, if $n$ is even, also for $k = n/2+1$) and $\chi_n(k) \triangleq 0$ otherwise. $b_1, \ldots, b_n$ are defined as follows: $\{b_k | 1 \leq k \leq n/2 + 1\}$ are independent RVs. If $\chi_n(k) = 1$ $b_k$ is uniformly distributed over $\{-1, 1\}$ (Rademacher), otherwise $b_k$ is uniformly distributed over the complex unit circle. For $n/2 + 1 < k \leq n$, $b_k = \overline{b}_{\langle n+2-k \rangle_n}$.

*Theorem 2 (distribution of RC mixture components)* [2]: With RC, for $1 \leq j, h \leq n$, we have the three following results:

(a) The distribution of $z_j$ is zero mean, symmetric, and

---

[2] A weaker version of part (b), namely $\Pr\{|z_j| > t\sigma_j\} \leq 2\exp[-t^2/4]$, may be derived from an intermediate step in the proof of Lemma 3.1 in [8].

$$\mathrm{cov}\{z_j, z_h\} = m^{-1}\rho_n(j-h) \tag{28}$$

where $\rho_n(l)$ is the circular autocorrelation of $\mathbf{x}$, that is, $\rho_n(l) \triangleq \sum_{k=1}^{n} x_k x_{\langle k+l\rangle_n}$, $|l| < n$. In particular,

$$\sigma_j^2 = \mathrm{cov}\{z_j, z_j\} = m^{-1}\|\mathbf{x}\|_2^2. \tag{29}$$

(b) If $\mathbf{x} \neq \mathbf{0}$ then

$$\Pr\{|z_j| > t\sigma_j\} \leq 2\exp\left[-t^2 \max\{0.25, \xi(t/\tau)\}\right] \tag{30}$$

where $\tau \triangleq \|\mathbf{x}\|_2 / \max_{1 \leq k \leq n}\left[(2 - \chi_n(k))|(F\mathbf{x})_k|\right]$ and

$$\xi(u) \triangleq u^{-2}\left[(1+u)\ln(1+u) - u\right] \tag{31}$$

(c) For any fixed integer $r > 0$, a sequence $\{\mathbf{u}^{(n)}\}$ as defined by (12) is AMN if condition (14) is satisfied and

$$\sup_n \|F^{(n)}\mathbf{x}^{(n)}\|_\infty < \infty \quad \square. \tag{32}$$

The parallels to Theorem 1 are apparent if we use (27) to view RC as complex variant of LR-SRM, with $F\mathbf{x}$ being the signal. Here too, $\mu_y = 0$ and $\sigma_y^2 = m^{-1}\|\mathbf{x}\|_2^2 = m^{-1}\|F\mathbf{x}\|_2^2$. Let $W = F^*$. Since correlations correspond to pointwise multiplication in the DFT domain, and since $\sqrt{n}\bar{f}_{jk}\bar{f}_{hk} = \bar{f}_{\langle j+1-h\rangle_n}$, we get from (28):

$$\mathrm{cov}\{z_j, z_h\} = m^{-1}\rho_n(j-h) = \frac{\sqrt{n}}{m}\mathbf{w}_{\langle j+1-h\rangle_n}\left((F\mathbf{x}) \circ \overline{(F\mathbf{x})}\right) = \frac{n}{m}\left(\mathbf{w}_j \circ \bar{\mathbf{w}}_h\right)\left((F\mathbf{x}) \circ \overline{(F\mathbf{x})}\right)$$

Therefore, except for the added complex conjugation, (28) is identical to (19) with $\mathbf{x}$ replaced by $F\mathbf{x}$.

The covariance of $\mathbf{z}$ can be computed efficiently $F\mathbf{x}$ is computed during the measurements computation and $\rho_n(0), \ldots, \rho_n(n-1)$ can be obtained from the inverse DFT of $\left[|(F\mathbf{x})_1|^2, \ldots, |(F\mathbf{x})_n|^2\right]^T$.

The function $\xi$ defined by (31) is slowly decreasing, with $\xi(0) = 0.5$ and $\xi(4.115) = 0.25$ (Fig. 1), hence the bound in the right hand side of (30) is between $\exp(-t^2/2)$ and $\exp(-t^2/4)$. If $n$ is moderately large and the signal does not have much of its energy concentrated in a specific frequency, then in the range of interest for $t$, $\tau_j \gg t$ and $\xi(t/\tau_j) \approx 0.5$, thus the bound on the tail distribution here is very close to the corresponding bound (21) for LR-SRM. Thus, for LR-SRM (11) is satisfied with $\eta(t, \tau_j) = 0.5$ and for RC it is satisfied with $\eta(t, \tau_j) \approx 0.5$.

A conclusion from Theorem 1 was that if all entries of $W$ have equal magnitude, then $\sigma_1 = \ldots = \sigma_n$. This condition is satisfied for $W = F^*$ and here too $\sigma_1 = \ldots = \sigma_n$.

Finally we note that the conditions for AMN are essentially the same here and in Theorem 1. Since $\|\mathbf{w}_j^{(n)}\|_\infty = n^{-1/2}$, condition

(13), required in Theorem 1, is satisfied here. Condition (14) is required in both theorems and condition (32) is the equivalent of the requirement that the signal be bounded when considering $F^{(n)}\mathbf{x}^{(n)}$ as the signal. In the RC case the covariance matrices $\{C_n\}$ are Toeplitz, which makes checking condition (14) for particularly simple.

In the following, because of the similarities, when referring to LR-SRM we mean to include Random Convolution as well.

*C. Quantization and Coding of LR-SRM Measurements*

We first consider coding LR measurements without linear prediction. In this case, only univariate distribution information needs to be shared with the decoder. If all the entries of the transform matrix have the same magnitude, as in the WHT case and the RC case, then the mixture component variances are the same and the conditions for asymptotic normality are satisfied with well behaved signals, i.e. signals which satisfy (24). Therefore $\sigma_y$ is the only parameter that needs to be shared as side information and all measurements can be coded by the same coding scheme and optimized for $\mathcal{N}(0, \sigma_y^2)$. This can be a good approximation for other transforms with entries of unequal magnitude, if $\sigma_1, \ldots, \sigma_n$ are similar. If there are significant differences among $\sigma_1, \ldots, \sigma_n$, the distribution of the measurements, which is a mixture of $z_1 \ldots, z_n$, can be approximated by a mixture of a smaller number of Gaussians, with the weights and variances of the approximation being shared as side information. The coding scheme for $z_{c(k)}$ is then optimized for the Guassian which represents it in the approximate mixture.

For the DCT and DFT, $\sigma_1, \ldots, \sigma_n$ may be approximated by some $\hat{\sigma}_1, \ldots, \hat{\sigma}_n$, using (23) and a parametric estimate of $W(\mathbf{x} \circ \mathbf{x})$. In this case it is possible to encode each measurement $y_k, 1 \leq k \leq m$ with a different coding scheme matched to $z_{c(k)}$, by assuming a distribution of $\mathcal{N}(0, \hat{\sigma}^2_{c(k)})$.

The above approaches, however, ignore the correlation among the mixture components, as shown by (19) and (28). Linear prediction can significantly reduce the data rate of the quantized measurements as outlined in Appendix A. The key to that approach is the finding of an approximation for $\text{cov}(z_j, z_h)$, $1 \leq j, h \leq q$, which can be shared with the decoder as side information. For LR-SRMs with the WHT, DFT and DCT this can be done using (22), (23), as explained in Sec. IV.A above. For RC, this can be done even more effectively because the covariance matrix is a Toeplitz matrix, hence the number of distinct matrix entries that need to be approximated is $n$ rather than $n(n+1)/2$. Furthermore, these distinct $n$ entries are the circular autocorrelation sequence of $\mathbf{x}$, which, for a typical media signal, can be represented by various alternative forms which are suitable for transmission and quantization [18]. If the conditions for asymptotic normality of the mixture components hold and $n$ is large enough, then the residuals are distributed normally and their estimated variance is given by (38), hence a coding scheme can be optimally adapted for each residual.

## V. GLOBAL RANDOMIZATION

Let $\pi : \mathbb{Z}_n \to \mathbb{Z}_n$ be a uniformly distributed random permutation of $\{1,\ldots,n\}$. Global randomization involves reordering of the columns of $W$ according to $\pi$, or equivalently, reordering of the entries of the input vector according to $\pi^{-1}$. A GR-SRM is a SRM as defined by (5) whose randomizer $R$ has in each row and each column exactly one non-zero element with a value of 1, and is given by $R_{jk} \triangleq \delta_{j,\pi(k)}$. Therefore,

$$z_j = \sqrt{n/m} \sum_{k=1}^{n} w_{j\pi(k)} x_k = \sqrt{n/m} \sum_{k=1}^{n} w_{jk} x_{\pi^{-1}(k)}, \quad 1 \le j \le n. \tag{33}$$

In this section, a bar over a vector or matrix symbol indicates the mean of the entries, e.g. $\bar{\mathbf{x}} \triangleq n^{-1} \mathbf{1}^T \mathbf{x}$, Let $T = I - n^{-1} \mathbf{1}_n \mathbf{1}_n^T$ be the linear operator which subtracts the mean from a signal. We can represent the signal as $\mathbf{x} = \bar{x}\mathbf{1} + T\mathbf{x}$. Clearly $\bar{x}\mathbf{1}$ is invariant under permutations, thus only the non-constant component of the signal is randomized.

### A. Measurements' Distribution with GR-SRM

With GR, any two mixture components, $z_j, z_h$, $1 \le j, h \le n$, have the same distribution if the entries of $\mathbf{w}_j, \mathbf{w}_h$ are a permutation of each other. Therefore, the number of distinct mixture components and their respective weights are determined solely by the transform $W$. For the WHT there are only two distinct mixture components: $\{z_1\}$ and $\{z_2, \ldots, z_n\}$, corresponding, respectively, to the constant $\mathbf{w}_1$, and to $\{\mathbf{w}_2, \ldots, \mathbf{w}_n\}$, each of which contains an equal number of $+n^{-1/2}$ and $-n^{-1/2}$ entries. For the DCT and DFT, the number of distinct mixture components is the number of divisors of $n$. In particular, if $n$ is a power of 2 there are $1 + \log_2 n$ distinct mixture components. If $W = W' \otimes W''$ the number of mixture components for $W$ does not exceed the product of the number of mixture components for $W'$ and $W''$. Thus, in these common examples, the number of distinct mixture components in the measurements' distribution is much smaller than $n$.

*Theorem 3 (distribution of GR-SRM mixture components)*[3]: With global randomization, for $1 \le j, h \le n$:

(a)

$$E\{z_j\} = n^{3/2} m^{-1/2} \bar{\mathbf{w}}_j \bar{\mathbf{x}}, \tag{34}$$

$$\text{cov}\{z_j, z_h\} = \frac{n}{m(n-1)} \left( \left(T\mathbf{w}_j^T\right)^T T\mathbf{w}_h^T \right) \|T\mathbf{x}\|_2^2 = \frac{n}{m(n-1)} \left( \delta_{jh} - n\bar{\mathbf{w}}_j \bar{\mathbf{w}}_h \right) \left( \|\mathbf{x}\|_2^2 - n\bar{\mathbf{x}}^2 \right) \tag{35}$$

---

[3] Part (b), with the assumption that $\bar{\mathbf{w}}_j = 0$, may be derived from an intermediate step in the proof of Theorem III.5 in [7]. The univariate case ($r=1$) of part (c) appears as Theorem III.1 in [7] with (13) and (15) replaced by (16) and (17) and the assumptions: $\bar{\mathbf{w}}_{j_n(1)}^{(n)} = 0, 1 \le j \le n$ and $\sqrt{n} \bar{\mathbf{x}}^{(n)} / \|\mathbf{x}^{(n)}\|_2 \to 0$.

$$\sigma_j^2 = \mathrm{cov}\{z_j, z_j\} = \frac{n}{m(n-1)} \|T\mathbf{w}_j^T\|_2^2 \|T\mathbf{x}\|_2^2 = \frac{n}{m(n-1)}(1 - n\bar{\mathbf{w}}_j^2)(\|\mathbf{x}\|_2^2 - n\bar{\mathbf{x}}^2) \tag{36}$$

(b) If $\sigma_j > 0$ then

$$\Pr\{|z_j - \mu_j| > t\sigma_j\} \leq 2\exp(-t^2 \eta(t, \tau_j)), \quad t \geq 0$$

where $\eta(t, \tau_j) \triangleq \tau_j^2 / (8n + 4t\tau_j)$, and

$$\tau_j = (n-1)^{-1/2} \frac{\|T\mathbf{w}_j^T\|_2 \, \|T\mathbf{x}\|_2}{\|T\mathbf{w}_j^T\|_\infty \, \|T\mathbf{x}\|_\infty} \leq \frac{n}{\sqrt{n-1}}. \tag{37}$$

(c) For any integer $r > 0$ a sequence $\{\mathbf{u}^{(n)}\}$ as defined by (12) is AMN if conditions (13) and (14) are satisfied □.

*Corollary:* With global randomization, for $1 \leq k \leq m$:

$$E\{y\} = n^{3/2} m^{-1/2} \bar{W}\bar{\mathbf{x}},$$
$$\sigma_y^2 \triangleq \mathrm{var}\{y\} = m^{-1}\left[\|\mathbf{x}\|_2^2 - n^3(\bar{\mathbf{x}}\bar{W})^2\right]$$

Unlike the LR case, Theorem 3 cleanly separates the signal and the transform by representing all significant parameters as products of functions of only $\mathbf{x}$ or only $W$. In addition, for a given $W$, the means and covariance structure of $z_1, \ldots, z_n$ are fully defined by $\bar{\mathbf{w}}_1, \ldots, \bar{\mathbf{w}}_n$, up to scaling factors determined by $\|\mathbf{x}\|_2$ and $\bar{\mathbf{x}}$. In particular, for any $1 \leq j, h \leq n$, $\sigma_j = \sigma_h$ if $\bar{\mathbf{w}}_j = \bar{\mathbf{w}}_h$.

Using the inequality in (37), we get

$$\tau_j^2/8n \geq \eta(t, \tau_j) \geq \tau_j^2/8n(1 + t/2\sqrt{n-1}).$$

Let $\tau_j = \alpha_\mathbf{x} \beta_j (n/(n-1))^{-1/2}$, where $\alpha_\mathbf{x} \triangleq n^{-1/2} \|T\mathbf{x}\|_2 / \|T\mathbf{x}\|_\infty \leq 1$, $\beta_j \triangleq n^{-1/2} \|T\mathbf{w}_j^T\|_2 / \|T\mathbf{w}_j^T\|_\infty \leq 1$. For moderately large $n$, $\sqrt{n/(n-1)} \approx 1$ and, in the range of interest for $t$, $t/2\sqrt{n-1} \approx 0$ hence $\eta(t, \tau_j) \approx \alpha_\mathbf{x}^2 \beta_j^2 / 8 \leq 1/8$. This bound on the tail distribution is singficantly weaker than the corresponding bound for local randomization, where $\eta(t, \tau_j) \approx 1/2$. The parameters $\alpha_\mathbf{x}$ and $\beta_j$, $j = 1, \ldots, n$ characterize the signal and the transform, respectively. If either one of them is small, this bound becomes weak. In the examples we considered $\beta_j$ also is not very small: if $j > 1$, $\beta_j^2 = 1$ for WHT and $\beta_j^2 = 1/2$ for DCT and DFT. For $W \triangleq W' \otimes W''$ let $\mathbf{w}'_{j'}, \mathbf{w}''_{j''}$ be rows of $W', W''$ respectively, with corresponding constants $\beta'_{j'}, \beta''_{j''}$ such that $\bar{\mathbf{w}}'_{j'} = \bar{\mathbf{w}}''_{j''} = 0$. Let $\mathbf{w}_j = \mathbf{w}'_{j'} \otimes \mathbf{w}''_{j''}$ then $\bar{\mathbf{w}}_j = 0$ and $\beta_j = \beta'_{j'} \beta''_{j''}$. The other parameter, $\alpha_\mathbf{x}$, becomes small if $T\mathbf{x}$ is compressible or sparse; for example, if $\mathbf{x}$ is a pulse train with period $d > 1$, then $\alpha_\mathbf{x}^2 = (d-1)^{-1}$.

## B. Quantization and Coding of GR-SRM Measurements

Mean and mean subtraction play a special role in GR-SRM because GR only randomizes the non-constant part of the signal.

Fortunately, in many practical transforms there is only one row $\mathbf{w}_j, 1 \leq j \leq n$ such that $\bar{\mathbf{w}}_j \neq 0$, say, $j=1$. This is true for the WHT, DCT and DFT, and if it is satisfied by $W', W''$ it is also satisfied by $W' \otimes W''$. For such transform matrices, $\mathbf{w}_1$ is constant (since $\{\mathbf{w}_2, \ldots, \mathbf{w}_n\}$ is a basis for the subspace of zero-mean vectors) and the means and covariances in (34), (35) get the form:

$$Ez_j = \delta_{1j} nm^{-1/2} \bar{\mathbf{x}},$$

$$\text{cov}(z_j, z_h) = \delta_{jh}(1-\delta_{1j})\left(\|\mathbf{x}\|_2^2 - n\bar{\mathbf{x}}^2\right) n/(m(n-1))$$

hence $z_2, \ldots, z_n$ are uncorrelated and have the same means and variances. Therefore, no redundancy can be removed by linear prediction. If in addition

$$\lim_{n \to \infty} \max_{1 < j \leq n} \|\mathbf{w}_j^{(n)}\|_\infty = 0,$$

$$\liminf_{n \to \infty} n^{-1} \|\mathbf{x}^{(n)} - \bar{\mathbf{x}}^{(n)} \mathbf{1}_n\|_2^2 > 0,$$

then condtions (13) and (14) are satisfied (for (14) this follows from $z_2, \ldots, z_n$ being uncorrelated), hence $z_2, \ldots, z_n$ are AMN, they asymptoically independent. Therefore, they can be quantized with the same quantizer, designed for $\mathcal{N}(0, \sigma_2^2) \approx \mathcal{N}\left(0, \left(\|\mathbf{x}\|_2^2 - n\bar{\mathbf{x}}^2\right)/m\right)$.

If $z_1$ is selected as a measurement, it cannot be quantized in the same way. However, in practice, this is not a problem because usually it is not necessary to quantize and transmit $z_1$ at all. Since $\bar{\mathbf{x}}$ and $\|\mathbf{x}\|_2$ determine the means and covarinces of $z_1, \ldots, z_n$, they are usually shared with the decoder as side information. Therefore, the decoder can calculate $z_1 = nm^{-1/2}\bar{\mathbf{x}}$ from the side information.

## VI. Normality Testing

We checked the normality of $z_j, j > 1$ for LR-SRM, GR-SRM and RST, each with the WHT, DCT and DFT. In the LR cases, the signal mean $\bar{\mathbf{x}}$ was subtracted from the signal prior to multiplying the signal by the sensing matrix. The test material was comprised of 160 video signals, each consisting of 12 frames of $88 \times 72$ 8-bit pixels. The 76032 pixels in each signal were zero padded to $n = 2^{17}$. In each test case, each signal vector was multiplied by 8 matrices of a specified type, which differed only in the random number generator seeds used to create the random matrices $S$ and $R$. $m$ was set to 76032, thus after removing the $z_1$ entries we had $8 \times 76031 = 608248$ measurements per signal. The mean and variance of those measurements were estimated and the measurements were normalized to zero mean and unit variance. If this set of measurements is a sample of a standard normal distribution, its quantile-quantile (Q-Q) plot [33] should be linear with a slope of 1. The Q-Q plots of the normalized measurements were computed for each of the signals and the 160 Q-Q plots were overlaid in one graph.

The graphs for each test case are shown in Fig. 2. In all cases of local and global randomization, the curves appear to be very close to linear, which indicates that the sample distribution is quite close to normal. Moderate deviations from the straight line appear above 3.5 standard deviations and may be explained by the scarcity of the data points in those regions. On the other hand, for RSTs the measurements distribution is far from normal and the shape of the Q-Q plots indicates that these distributions have much heavier tails than normal distributions with the same variance. The wide spread among the Q-Q plots in RSTs shows that the shape of the measurements' distribution is highly dependent on the input signal.

According to the analysis of the LR case (Sec. IV), asymptotic normality is guaranteed for measurements generated by WHT, but signal dependent deviations from normality may appear in measurements generated by DCT or DFT. Yet, these deviations are hardly noticeable in Fig. 2. Fig. 3 shows an enlarged section (the range of [1,1.5]) of the same graphs. While the graphs of the WHT case retain its clean, linear shape, the graph of the DCT case is a wide stripe with irregular boundaries, indicating that its underlying individual Q-Q plots deviated slightly from the linear shape expected for a normal distribution, and that these deviations are signal dependent, making each Q-Q plot different. The Q-Q graph for the DFT case is as clean as the one of the WHT, which confirms our expectation that the constraint (25) will cause $\sigma_1^2,\ldots,\sigma_n^2$ to be close to $\sigma_y^2$.

## VII. Discussion

SRMs were introduced in order to make transform based sensing matrices universal. Our work shows that the signal randomization of SRMs makes either the measurements, or the mixture components they are derived from, suitable for quantization by reducing the tail probability of their distributions and making them approximately multivariate normal, under a wide variety of conditions. In contrast, our simulations show that it would be quite difficult to efficiently quantize measurements generated by RSTs because of the wide spread of their distributions and because of the complex dependence of the distribution on the input signal.

Random Convolution [8] is another method which was proposed independently for the same purpose after SRMs were introduced [6]. We showed that it can be viewed a variant of LR-SRM, with similar statistical properties of measurements.

If we assume that $n$ is sufficiently large, and, for GR, that $\mathbf{w}_1$ is constant and $z_1$ is not selected (it may be available to the decoder as side information), then the measurements in both GR and LR are approximately IID and have similar variances. However, the corresponding mixture components behave quite differently. In both cases they are approximately multivariate normal, but in the GR case the mixture components are also approximately IID, while in the LR case they can be highly correlated. A possible reason for this difference is that while both LR and GR generate an uncorrelated random vector $\tilde{\mathbf{x}} = R\mathbf{x}$,

GR does a more thorough randomization job: with GR the sequence $\{\tilde{x}_k^2\}$ is also uncorrelated, whereas with LR $\tilde{x}_k^2 = x_k^2, 1 \leq k \leq n$ (the same is true for higher even-order moments).

Within the LR family, there are differences in the univariate distribution of mixture components between transform matrices with equal magnitude entries, such as the WHT and RC, and those with different magnitude entries, such as the DCT and DFT (in its real version). With the former, $z_1, \ldots, z_n$ have identical variance, while with the latter the variances are different, hence the measurements distribution is not asymptotically normal. As our normality testing simulation showed, the deviations from normality vary depending on the type of transform and may be small enough to be ignored.

Based on these findings we proposed several methods to improve the quantization and coding of compressive measurements by using signal-specific parameters. These parameters are computed by the encoder and conveyed to the decoder as side information, enabling the decoder to adapt the dequantizer and channel decoder so that they match the quantizer and channel encoder. The applicability of these methods and the amount of side information which needs be shared depend on the flavor of SRM used and the properties of the input signal. For GR-SRM, the only information which needs to be shared is $\bar{\mathbf{x}}$ and $\|\mathbf{x}\|_2$, and the achieveable bit rate is the one achievalbe for a guassian memoryless source with variance $\sigma_y^2$. On the other hand, for LR-SRM it is possible to share a parameteric model of the covariance of the mixture components and then significantly reduce the coded measuremets bit rate by removing reducndancy, e.g. by the use of linear prediction.

The differences in measurements distribution may lead to design preferences for local or global randomization, or for a particular transform. However, the designer's choice may be restricted by the application, especially if the measurements are generated in the analog domain. For example, LR may be impractical in a single pixel camera [32] because it requires negative optical amplification. Similarly, the transform may be dictated by the nature of the application, and may not necessarily be one of the common transforms which we have used as examples.

APPENDIX A: DECORRELATION BY LINEAR PREDICTION

Suppose we have a parametric model approximation to $\operatorname{cov}\{z_j, z_i\}$, $1 \leq i, j \leq n$, and the parameters of the model are shared with the decoder as side information. Let $r > 0$ be a fixed integer and for simplicity assume $m$ is an integer multiple of $r$. Let $z_{c(1)}, \ldots, z_{c(m)}$ be grouped in $m/r$ sets $\{z_{1p}, \ldots, z_{rp}\}$, $1 \leq p \leq m/r$ such that the $r$ members of each set are significantly correlated. For $1 \leq q \leq r$, $1 \leq p \leq n/r$, let $a_{h,q,p}, 1 \leq h < q$ be the linear prediction coefficients which minimize $\operatorname{var}\{u_{qp}\}$, where the residual $u_{qp}$ is defined by

$$u_{qp} \triangleq \left(z_{qp} - E\{z_{qp}\}\right) - \sum_{h=1}^{q-1} a_{h,q,p}\left(z_{hp} - E\{z_{hp}\}\right)$$

As is well known, $a_{h,q,p}, 1 \leq h < q$ are the solution of a set of $q-1$ linear equations whose coefficients are $\text{cov}(z_{kp}, z_{lp})$, $1 \leq k, l \leq q$. $\{u_{1p}, \ldots, u_{qp}\}$ are zero mean, uncorrelated and

$$\text{var}\{u_{qp}\} = \text{var}\{z_{qp}\} - \sum_{h=1}^{q-1} a_{h,q,p} \text{cov}(z_{hp}, z_{qp}). \tag{38}$$

If $\{z_{1p}, \ldots, z_{rp}\}$ are multivariate normal, then $\{u_{1p}, \ldots, u_{qp}\}$ are multivariate normal and independent. In our case the linear prediction coefficients and $\text{var}\{u_{qp}\}$ are computed using the parametric approximation of the covariance. Let $Q_{qp}$ be a quantizer optimized for the residual $u_{qp}$. Quantization is performed recursively for $q = 1, \ldots, r$ by

$$\hat{u}_{qp} = Q_{qp}\left((z_{qp} - E\{z_{qp}\}) - \sum_{h=1}^{q-1} a_{h,q,p}(\hat{z}_{hp} - E\{z_{hp}\})\right)$$

$$\hat{z}_{qp} = \hat{u}_{qp} + E\{z_{qp}\} + \sum_{h=1}^{q-1} a_{h,q,p}(\hat{z}_{hp} - E\{z_{hp}\}). \tag{39}$$

The code words $\{\hat{u}_{qp} \mid 1 \leq q \leq r, 1 \leq p \leq n/r\}$ are channel-coded and transmitted as side information. The decoder uses the approximate covariance model to group $z_{c(1)}, \ldots, z_{c(m)}$ into $r$-sized sets and to estimate the prediction coefficients in the same way sets as the encoder. Then, the decoder uses (39) to compute the dequantized measurements $\{\hat{z}_{1p}, \ldots, \hat{z}_{rp}\}$.

## APPENDIX B: PROOFS

*Proof of Lemma 1*: Since $|C_0^{(n)}| = n^m$, $|C_1^{(n)}| = n(n-1)\ldots(n-m+1)$, and $m = m(n) = o(\sqrt{n})$:

$$0 \geq \log \frac{|C_1^{(n)}|}{|C_0^{(n)}|} = \sum_{k=0}^{m-1} \log\left(1 - \frac{k}{n}\right) \geq -\sum_{k=0}^{n} \frac{k}{n} = \frac{-m(m-1)}{2n} \xrightarrow[n \to \infty]{} 0$$

hence $|C_1^{(n)}|/|C_0^{(n)}| \xrightarrow[n \to \infty]{} 1$. For any $\mathbf{g} \in C_0^{(n)}$ let $G_\mathbf{g}^{(n)}$ be the multivariate distribution function of $[z_{g_1}^{(n)}, \ldots, z_{g_{m(n)}}^{(n)}]^T$. Then

$$F_i^{(n)}(\mathbf{h}^{(n)}) = |C_i^{(n)}|^{-1} \sum_{\mathbf{g} \in C_i^{(n)}} G_\mathbf{g}^{(n)}(\mathbf{h}^{(n)}), \quad i = 0, 1$$

Therefore,

$$F_1^{(n)}(\mathbf{h}^{(n)}) = |C_1^{(n)}|^{-1} \left[\sum_{\mathbf{g} \in C_0^{(n)}} G_\mathbf{g}^{(n)}(\mathbf{h}^{(n)}) - \sum_{\mathbf{g} \in C_0^{(n)} - C_1^{(n)}} G_\mathbf{g}^{(n)}(\mathbf{h}^{(n)})\right] = |C_0^{(n)}||C_1^{(n)}|^{-1} F_0^{(n)}(\mathbf{h}^{(n)}) - |C_1^{(n)}|^{-1} \sum_{\mathbf{g} \in C_0^{(n)} - C_1^{(n)}} G_\mathbf{g}^{(n)}(\mathbf{h}^{(n)})$$

$$\frac{F_1^{(n)}(\mathbf{h}^{(n)})}{F_0^{(n)}(\mathbf{h}^{(n)})} = \frac{|C_0^{(n)}|}{|C_1^{(n)}|} - \frac{|C_1^{(n)}|^{-1}}{F_0^{(n)}(\mathbf{h}^{(n)})} \sum_{\mathbf{g} \in C_0^{(n)} - C_1^{(n)}} G_\mathbf{g}(\mathbf{h}^{(n)})$$

The second term on the right hand side vanishes as $n \to \infty$ because $\liminf_{n \to \infty} F_0^{(n)}(\mathbf{h}^{(n)}) > 0$ and

$$0 \leq |C_1^{(n)}|^{-1} \sum_{\mathbf{g} \in C_0^{(n)} - C_1^{(n)}} G_\mathbf{g}(\mathbf{h}^{(n)}) \leq |C_1^{(n)}|^{-1}(|C_0^{(n)}| - |C_1^{(n)}|) \to 0.$$

Therefore

$$\lim_{n\to\infty} F_1^{(n)}\left(\mathbf{h}^{(n)}\right)\Big/ F_0^{(n)}\left(\mathbf{h}^{(n)}\right) = \lim_{n\to\infty} \left|C_0^{(n)}\right|\Big/\left|C_1^{(n)}\right| = 1 \quad \square.$$

*Proof of Lemma 2*: We need to show that for any $\varepsilon > 0$ and $\mathbf{v} \in \mathbb{R}^r$ there is $n_0(\varepsilon, \mathbf{v})$ such that if $n \geq n_0(\varepsilon, \mathbf{v})$ and $\{\mathbf{u}^{(n)}\}$ satisfies (12) with $\{j_n(1),\ldots,j_n(r)\} \subseteq J^{(n)}$, then $\left|F(\mathbf{v}) - F_{\mathbf{u}^{(n)}}(\mathbf{v})\right| \leq \varepsilon$, where $F(\mathbf{v})$, $F_{\mathbf{u}^{(n)}}(\mathbf{v})$ denote the probability distributions of $\mathcal{N}(\mathbf{0}_r, I_r)$ and of $\{\mathbf{u}^{(n)}\}$, respectively. Suppose this was not true. Then we could create a sequence $\{\mathbf{u}^{*(n)}\}$ which, for sufficiently large $n$, satisfies (12) and for which $\{j_n(1),\ldots,j_n(r)\} \subseteq J^{(n)}$, but $\limsup_n \left|F(\mathbf{v}) - F_{\mathbf{u}^{*(n)}}(\mathbf{v})\right| \geq \varepsilon$, in contradiction to the lemma's assumption that $\{\mathbf{u}^{*(n)}\}$ is AMN $\square$.

The proofs of theorems 1-3 rely on the following theorems and on Lemma 3 below:

*Hoeffding Concentration Inequality* [34]: Let $\{X_k | k=1,\ldots,n\}$ be independent RVs such that $\Pr\{a_k \leq |X_k| \leq b_k\} = 1$. Then

$$\Pr\left\{\left|\sum_{k=1}^n (X_k - E\{X_k\})\right| \geq \varepsilon\right\} \leq 2\exp\left[-2\varepsilon^2 \Big/ \sum_{k=1}^n |b_k - a_k|\right] \quad \square.$$

*Bennett's Concentration Inequality* [35]: Let $\{X_k | k=1,\ldots,n\}$ be independent RVs such that $\Pr\{|X_k| \leq D\} = 1$. Let $v^2 \triangleq \sum_{k=1}^n E\{X_k^2\}$. For any $\varepsilon \geq 0$,

$$\Pr\left\{\left|\sum_{k=1}^n (X_k - E\{X_k\})\right| \geq \varepsilon\right\} \leq \exp\left[-\frac{v^2}{D^2}\theta\left(\frac{\varepsilon D}{v^2}\right)\right]$$

where

$$\theta(u) \triangleq (1+u)\log(1+u) - u \quad \square.$$

*Lyapunov Central Limit Theorem for Triangular Arrays* [36]: Let $X_{nk}, 1 \leq k \leq p_n, n=1,2,\ldots$ be RVs, $E\{X_{nk}\} = \mu_{nk}$, $\text{var}\{X_{nk}\} = \sigma_{nk}^2 < \infty$, such that for each $n$, $X_{n1},\ldots X_{np_n}$ are independent. Let $s_n^2 \triangleq \sum_{k=1}^{p_n} \sigma_{nk}^2$. If

$$\lim_{n\to\infty} s_n^{-(2+\delta)} \sum_{k=1}^{p_n} E\left\{|X_{nk} - \mu_{nk}|^{2+\delta}\right\} = 0 \tag{40}$$

for some $\delta > 0$, then:

$$s_n^{-1} \sum_{k=1}^{p_n} (X_{nk} - \mu_{nk}) \xrightarrow[n\to\infty]{d} \mathcal{N}(0,1) \quad \square.$$

*Chatterjee Concentration Inequality* [37]: Let $a_{ij} \in [0,1], 1 \leq i, j \leq n$ and let $S \triangleq \sum_{k=1}^n a_{k,\pi(k)}$, where $\pi$ is a uniformly distributed, random permutation of $\{1,\ldots,n\}$. For $t \geq 0$

$$\Pr\{|S - E\{S\}| \geq t\} \leq 2\exp\left[-t^2 \big/ (4E\{S\} + 2t)\right] \quad \square.$$

*Combinatorial Central Limit Theorem* [38]: For $n=1,2,\ldots$ let $\{a_n(i)\}$, $\{b_n(i)\}$, $1 \leq i \leq n$ be non-constant sequences with means $\bar{a}_n, \bar{b}_n$, respectively. Let $\pi_n$ be a uniformly distributed permutation of $\{1,\ldots,n\}$. Define $S_n \triangleq \sum_{i=1}^{n} a_n(i) b_n(\pi_n(i))$. If

$$\lim_{n \to \infty} n \frac{\max_{1 \leq i \leq n}(a_n(i) - \bar{a}_n)^2 \max_{1 \leq i \leq n}(b_n(i) - \bar{b}_n)^2}{\sum_{i=1}^{n}(a_n(i) - \bar{a}_n)^2 \sum_{i=1}^{n}(b_n(i) - \bar{b}_n)^2} = 0 \tag{41}$$

then $S_n$ is asymptotically normally distributed $\square$.

*Lemma 3*: If $\{\mathbf{u}^{(n)}\}$ is defined by (12), condition (14) is satisfied, and $(m/n)\{\text{var}\{u_k^{(n)}\} \mid 1 \leq k \leq r, n=1,\ldots\}$ is bounded then for sufficiently large $n$ there is $\varepsilon > 0$ such that

$$\forall \boldsymbol{\alpha} \in \mathbb{R}^r : (m/n) \boldsymbol{\alpha}^T \text{cov}(\mathbf{u}^{(n)}) \boldsymbol{\alpha} \geq \varepsilon \|\boldsymbol{\alpha}\|_2^2 \quad \square.$$

*Proof*: Let $D$ be such that $(m/n) \text{var}\{u_k^{(n)}\} \leq D, 1 \leq k \leq r, n=1,\ldots$ Then the magnitude of each entry of $(m/n)\text{cov}(\mathbf{u}^{(n)})$ is bound by $D$ and therefore $\|(m/n)\text{cov}(\mathbf{u}^{(n)})\|_2 \leq rD$. Let $S \triangleq \{\boldsymbol{\alpha} \in \mathbb{R}^r \mid \|\boldsymbol{\alpha}\|_2 = 1\}$ be the unit sphere in $\mathbb{R}^r$. For $\boldsymbol{\alpha} \in \mathbb{R}^r$ let

$f_n(\boldsymbol{\alpha}) \triangleq (m/n) \boldsymbol{\alpha}^T \text{cov}(\mathbf{u}^{(n)}) \boldsymbol{\alpha}, n=1,\ldots$, and $f(\boldsymbol{\alpha}) \triangleq \liminf_n f_n(\boldsymbol{\alpha})$. If $\boldsymbol{\alpha}_1, \boldsymbol{\alpha}_2 \in S$ then $\|\boldsymbol{\alpha}_1 + \boldsymbol{\alpha}_2\|_2 \leq 2$ and

$$|f_n(\boldsymbol{\alpha}_1) - f_n(\boldsymbol{\alpha}_2)| = (m/n)|(\boldsymbol{\alpha}_1 + \boldsymbol{\alpha}_2)^T \text{cov}(\mathbf{u}^{(n)})(\boldsymbol{\alpha}_1 - \boldsymbol{\alpha}_2)| \leq 2D \|\boldsymbol{\alpha}_1 - \boldsymbol{\alpha}_2\|_2.$$

Therefore, $f_n$ is continuous and since $S$ is compact, $f_n$ has a minimum on $S$ at some $\boldsymbol{\alpha}^{(n)} \in S$. The sequence $\{(f_n, \boldsymbol{\alpha}^{(n)})\}_n$ has a subsequence $\{(f_{n(h)}, \boldsymbol{\alpha}^{(n(h))})\}_h$ such that $f_{n(h)}(\boldsymbol{\alpha}^{(n(h))}) \to \liminf_n f_n(\boldsymbol{\alpha}^{(n)})$ and $\boldsymbol{\alpha}^{(n(h))} \to \boldsymbol{\alpha}^* \in S$. By (14)

$$0 < f(\boldsymbol{\alpha}^*) = \liminf_h f_{n(h)}(\boldsymbol{\alpha}^*) \leq \lim_h \left[ f_{n(h)}(\boldsymbol{\alpha}^{(n(h))}) + 2D \|\boldsymbol{\alpha}^{(n(h))} - \boldsymbol{\alpha}^*\|_2 \right] = \liminf_n f_n(\boldsymbol{\alpha}^{(n)})$$

Select $0 < \varepsilon < f(\boldsymbol{\alpha}^*)$. For sufficiently large $n$ and for any $\boldsymbol{\alpha} \in \mathbb{R}^r$, if $\boldsymbol{\alpha} \neq \mathbf{0}$ then $\|\boldsymbol{\alpha}\|_2^{-1} \boldsymbol{\alpha} \in S$. Therefore,

$$f_n(\boldsymbol{\alpha}) = \|\boldsymbol{\alpha}\|_2^2 f_n(\|\boldsymbol{\alpha}\|_2^{-1} \boldsymbol{\alpha}) \geq \|\boldsymbol{\alpha}\|_2^2 f_n(\boldsymbol{\alpha}_n) \geq \varepsilon \|\boldsymbol{\alpha}\|_2^2 \quad \square.$$

*Proof of Theorem 1*: (a) Let $X_{jk} \triangleq (n/m)^{1/2} w_{jk} x_k b_k$. Symmetry, (18), (19) and (20) hold because $z_j = \sum_{k=1}^{n} X_{jk}$ and $X_{jk}, k=1,\ldots,n$ are symmetric, independent, $E\{X_{jk}\} = 0$ and $E\{X_{jk} X_{hl}\} = (n/m) \delta_{kl} w_{jk} w_{hk} x_k^2$.

(b)

$|X_{jk}| \leq (n/m)^{1/2} |w_{jk}||x_k|$. Therefore, by Hoeffding concentration inequality and (20):

$$\Pr\{|z_j| > t\sigma_j\} \leq \exp\left[-2t^2 \sigma_j^2 \bigg/ \sum_{k=1}^{n} 4(n/m) w_{jk}^2 x_k^2 \right] = \exp[-t^2/2]$$

(c) Let $C_n \triangleq \text{cov}(\mathbf{u}^{(n)})$. By (20) $(m/n)\sigma_{jn}^2 \leq x_{\max}^2$ hence by lemma 3 that there is $\varepsilon > 0$ such that for sufficiently large $n$, $C_n$ is positive definite and for any $\boldsymbol{\alpha} \in \mathbb{R}^r$, $\boldsymbol{\alpha}^T C_n \boldsymbol{\alpha} \geq \varepsilon^2 (n/m) \|\boldsymbol{\alpha}\|_2^2$. For such $n$ define

$$\tilde{\boldsymbol{\alpha}}_n \triangleq C_n^{-1/2} \boldsymbol{\alpha}$$

$$\tilde{\mathbf{w}}_k^{(n)} \triangleq [w_{j(1),k}^{(n)}, \ldots, w_{j(r),k}^{(n)}]^T, \quad X_{nk} \triangleq \sqrt{n/m} x_k b_k \tilde{\boldsymbol{\alpha}}_n^T \tilde{\mathbf{w}}_k^{(n)}, \quad 1 \leq k \leq n$$

Then $X_{n1}, \ldots X_{nn}$ are independent, zero mean RVs and:

$$E\{X_{nk}^2\} = \tilde{\boldsymbol{\alpha}}_n^T \left[ (n/m) x_k^2 \tilde{\mathbf{w}}_k^{(n)} \left(\tilde{\mathbf{w}}_k^{(n)}\right)^T \right] \tilde{\boldsymbol{\alpha}}_n$$

$$s_n^2 \triangleq \sum_{k=1}^n E\{X_{nk}^2\} = \tilde{\boldsymbol{\alpha}}_n^T C_n \tilde{\boldsymbol{\alpha}}_n = \|\boldsymbol{\alpha}\|_2^2$$

Condition (40) of Lyapunov CLT for triangular arrays is satisfied because for any $\delta > 0$, using (13):

$$s_n^{-(2+\delta)} \sum_{k=1}^n E\{|X_{nk}|^{2+\delta}\} \leq s_n^{-(2+\delta)} \max_{1 \leq k \leq n} |X_{nk}|^\delta s_n^2 \leq s_n^{-\delta} \left[ x_{\max} \sqrt{n/m} \max_{1 \leq k \leq n} |\tilde{\mathbf{w}}_k^{(n)} \tilde{\boldsymbol{\alpha}}_n| \right]^\delta \leq$$

$$\|\boldsymbol{\alpha}\|_2^{-\delta} \left[ x_{\max} \sqrt{n/m} \|\tilde{\boldsymbol{\alpha}}_n\|_2 \max_{1 \leq k \leq n} \|\tilde{\mathbf{w}}_k^{(n)}\|_2 \right]^\delta \leq \|\boldsymbol{\alpha}\|_2^{-\delta} \left[ x_{\max} \varepsilon^{-1} \sqrt{\tilde{\boldsymbol{\alpha}}_n^T C_n^{-1} \tilde{\boldsymbol{\alpha}}_n} \right]^\delta r^{\delta/2} \max_{1 \leq k \leq n} \|\mathbf{w}_{j(k)}^{(n)}\|_\infty^\delta \leq \left( x_{\max} \varepsilon^{-1} \sqrt{r} \right)^\delta \max_{1 \leq k \leq n} \|\mathbf{w}_{j(k)}^{(n)}\|_\infty^\delta \xrightarrow{n \to \infty} 0.$$

Therefore $\boldsymbol{\alpha}^T C_n^{-1/2} \mathbf{u}^{(n)} = \tilde{\boldsymbol{\alpha}}_n^T \left[ z_{j(1)}^{(n)}, \ldots, z_{j(r)}^{(n)} \right]^T = \sum_{k=1}^n X_{nk} \xrightarrow{d} \mathcal{N}(0, \|\boldsymbol{\alpha}\|_2^2)$. Since this is true for all $\boldsymbol{\alpha} \in \mathbb{R}^r$, $\boldsymbol{\alpha} \neq \mathbf{0}$, then by Cramér-Wold Theorem [36], $C_n^{-1/2} \mathbf{u}^{(n)} \xrightarrow{d} \mathcal{N}(0, I_r)$ □.

*Proof of Theorem 2:* (a) Let $\beta_k \triangleq \arg(b_k)$, $k = 1, \ldots, n$. Then $\{\beta_k | 1 \leq k \leq (n+2/2)\}$ are independent and for $(n+2)/2 < k \leq n$, $\beta_k = 2\pi - \beta_{n+2-k}$. Each of $\beta_1 \ldots \beta_n$ is uniformly distributed. If $\chi_n(k) = 1$ the uniform distribution is over $\{0, \pi\}$, otherwise it is over $[0, 2\pi)$. Let $\mathbf{v} = [v_1, \ldots, v_n]^T = F\mathbf{x}$. For $1 \leq j, k \leq n$ let

$$\varphi_{jk} \triangleq \arg(v_k) + 2\pi n^{-1}(j-1)(k-1)$$

$$X_{jk} \triangleq m^{-1/2} (2 - \chi_n(k)) |v_k| \cos(\beta_k + \varphi_{jk})$$

Note that $\arg(v_k)$ and $\varphi_{jk}$ are multiples of $\pi$ if $\chi_n(k) = 1$. By (26) and the definition of $F^*$:

$$z_j = (n/m)^{1/2} \sum_{k=1}^n \bar{f}_{jk} b_k v_k = \sum_{1 \leq k \leq n/2+1} m^{-1/2} (2 - \chi_n(k)) \mathrm{Re}\{n^{1/2} \bar{f}_{jk} b_k v_k\} = \sum_{1 \leq k \leq n/2+1} X_{jk}$$

The distributions of $X_{jk}$, $1 \leq k \leq n/2+1$ are zero mean and symmetric, hence the distribution of $z_j$, $1 \leq j \leq n$ is also zero mean and symmetric. For any $1 \leq j, h \leq n$, $1 \leq k, l \leq n/2+1$

$$E\{\cos(\beta_k + \varphi_{jk}) \cos(\beta_l + \varphi_{hl})\} = \frac{\delta_{kl}}{2} E\{\cos(\varphi_{jk} - \varphi_{hk}) + \cos(2\beta_k + \varphi_{jk} + \varphi_{hk})\} = \frac{\delta_{kl}(1 + \chi_n(k))}{2} \cos(\varphi_{jk} - \varphi_{hk})$$

hence

$$E\{X_{jk} X_{hl}\} = m^{-1} \delta_{kl} (2 - \chi_n(k)) |v_k|^2 \cos(\varphi_{jk} - \varphi_{hk})$$

Let $\tilde{v}_k \triangleq |v_k|^2$, $1 \le k \le n$, $\tilde{\mathbf{v}} \triangleq [\tilde{v}_1, \ldots, \tilde{v}_n]^T$. By the DFT properties, $\rho_n(j) = n^{1/2}(F^*\tilde{\mathbf{v}})_{j+1}$. (28) is obtained from

$$E\{z_j z_h\} = n^{-1} \sum_{1 \le k \le n/2+1} E\{X_{jk} X_{hk}\} = m^{-1} \sum_{1 \le k \le n/2+1} (2 - \chi_n(k)) \tilde{v}_k \cos(\varphi_{jk} - \varphi_{hk}) = m^{-1} n^{1/2} \operatorname{Re}\{(F^*\tilde{\mathbf{v}})_{j-h+1}\} = m^{-1} \rho_n(j-h) \quad \text{and} \quad (29)$$

follows because $\rho(0) = \|\mathbf{x}\|_2^2$.

(b) $X_{jk}$, $1 \le k \le n/2+1$ are independent and $|X_{jk}| \le D_{jk} \triangleq m^{-1/2}(2 - \chi_n(k))|v_k|$.

$$\sum_{1 \le k \le n/2+1} (2D_{jk})^2 \le 8m^{-1} \sum_{k=1}^{n} |v_k|^2 = 8m^{-1} \|\mathbf{x}\|_2^2 = 8\sigma_j^2,$$

hence by Hoeffding concentration inequality:

$$\Pr\{|z_j| > t\sigma_j\} \le 2\exp\left[\frac{-2t^2\sigma_j^2}{\sum_{1 \le k \le n/2+1}(2D_{jk})^2}\right] \le 2\exp\left[\frac{-t^2}{4}\right].$$

If $\mathbf{x} \ne \mathbf{0}$ then $\sum_{1 \le k \le n/2+1} E\{X_{jk}^2\} = \sigma_j^2 = m^{-1}\rho^2(0) = m^{-1}\|\mathbf{x}\|_2^2 > 0$. Let

$$D \triangleq m^{-1/2} \max_{1 \le h \le n}\left[(2 - \chi_n(h))|(F\mathbf{x})_h|\right] \ge |X_{jk}|, \quad 1 \le j, k \le n$$

By Bennett concentration inequality, since $\tau = \sigma_j/D$:

$$\Pr\{|z_j| > t\sigma_j\} \le 2\exp\left[-\tau_j^2 \theta(t/\tau_j)\right] = 2\exp\left[-t^2 \xi(t/\tau_j)\right].$$

(c) Let $C_n \triangleq \operatorname{cov}(\mathbf{u}^{(n)})$. By (29) $(m/n)\sigma_{jn}^2 \le n^{-1}\|\mathbf{x}^{(n)}\|_2^2 \le x_{\max}^2$ hence by lemma 3 that there is $\varepsilon > 0$ such that for sufficiently large $n$, $C_n$ is positive definite and for $\boldsymbol{\alpha} \in \mathbb{R}^r$, $\boldsymbol{\alpha}^T C_n \boldsymbol{\alpha} \ge \varepsilon^2 (n/m)\|\boldsymbol{\alpha}\|_2^2$. For such $n$ and any $\boldsymbol{\alpha} \in \mathbb{R}^r$ define $\tilde{\boldsymbol{\alpha}}_n \triangleq C_n^{-1/2}\boldsymbol{\alpha}$, $[v_1^{(n)}, \ldots, v_n^{(n)}]^T \triangleq F^{(n)}\mathbf{x}^{(n)}$. For $1 \le k \le n/2+1$ define

$$\tilde{\mathbf{w}}_k^{(n)} \triangleq [\overline{f}_{j(n,1),k}^{(n)}, \ldots, \overline{f}_{j(n,r),k}^{(n)}]^T$$

$$\varphi_{nk} \triangleq \arg(v_k^{(n)} \tilde{\boldsymbol{\alpha}}_n^T \tilde{\mathbf{w}}_k^{(n)})$$

$$X_{nk} \triangleq (n/m)^{1/2}(2 - \chi_n(k))\operatorname{Re}\{b_k v_k^{(n)} \tilde{\boldsymbol{\alpha}}_n^T \tilde{\mathbf{w}}_k^{(n)}\} = (n/m)^{1/2}(2 - \chi_n(k))|v_k^{(n)} \tilde{\boldsymbol{\alpha}}_n^T \tilde{\mathbf{w}}_k^{(n)}|\cos(\beta_k + \varphi_{nk})$$

$\{X_{nk}, 1 \le k \le n/2+1\}$ are independent, zero mean, and:

$$E\{X_{nk}^2\} = (n/m)(2 - \chi_n(k))|\tilde{\boldsymbol{\alpha}}_n^T \tilde{\mathbf{w}}_k^{(n)}|^2 |v_k|^2 = (n^{1/2}/m)(2 - \chi_n(k))|\tilde{\boldsymbol{\alpha}}_n^T \tilde{\mathbf{w}}_k^{(n)}|^2 \sum_{l=0}^{n-1} \overline{f}_{k,\langle l\rangle_n}^{(n)} \rho(l),$$

$$|X_{nk}| \le 2v_{\max}\sqrt{n/m} \max_{1 \le k \le n/2+1}|\tilde{\boldsymbol{\alpha}}_n^T \tilde{\mathbf{w}}_k^{(n)}| \le 2v_{\max}\sqrt{n/m}\|\tilde{\boldsymbol{\alpha}}_n\|_2 \max_{1 \le k \le n}\|\tilde{\mathbf{w}}_k^{(n)}\|_2 \le 2v_{\max}\varepsilon^{-1}\sqrt{\tilde{\boldsymbol{\alpha}}_n^T C_n \tilde{\boldsymbol{\alpha}}_n}\sqrt{r/n} \le 2v_{\max}\varepsilon^{-1}\|\boldsymbol{\alpha}\|\sqrt{r/n} \xrightarrow[n \to \infty]{} 0$$

where $v_{\max} \triangleq \sup_n \|\mathbf{v}^{(h)}\|_\infty < \infty$ because of (32). Let

$$s_n^2 \triangleq \sum_{1 \le k \le n/2+1} E\{X_{nk}^2\} = (n^{1/2}/m) \sum_{k=1}^{n} |\tilde{\alpha}_n^T \tilde{\mathbf{w}}_k^{(n)}|^2 \sum_{l=0}^{n-1} \overline{f}_{k,\langle l \rangle_n}^{(n)} \rho(l) = (n^{1/2}/m) \sum_{l=0}^{n-1} \rho(l) \sum_{h,h'=1}^{r} \tilde{\alpha}_{nh} \tilde{\alpha}_{nh'} \sum_{k=1}^{n} f_{j(n,h),k}^{(n)} \overline{f}_{j(n,h'),k}^{(n)} \overline{f}_{k,\langle l \rangle_n}^{(n)} =$$

$$(n^{1/2}/m) \sum_{l=0}^{n-1} \rho(l) \sum_{h,h'=1}^{r} \tilde{\alpha}_{nh} \tilde{\alpha}_{nh'} n^{-1/2} \delta_{l,\langle j(n,h)-j(n,h') \rangle} = \tilde{\alpha}_n^T C_n \tilde{\alpha}_n = \|\boldsymbol{\alpha}\|_2^2$$

Lyapunov condition holds for $\{X_{nk} \mid 1 \le k \le n/2+1\}$ because

$$s_n^{-(2+\delta)} \sum_{k=1}^{n} E\{|X_{nk}|^{2+\delta}\} \le s_n^{-(2+\delta)} \max_{1 \le k \le n} |X_{nk}|^{\delta} s_n^2 = \|\boldsymbol{\alpha}\|_2^{-\delta} \max_{1 \le k \le n} |X_{nk}|^{\delta} \xrightarrow[n \to \infty]{} 0.$$

Therefore

$$\boldsymbol{\alpha}^T C_n^{-1/2} \mathbf{u}^{(n)} = \tilde{\boldsymbol{\alpha}}_n^T \mathbf{u}^{(n)} = \sum_{k=1}^{n} (n/m)^{1/2} v_k b_k \tilde{\boldsymbol{\alpha}}_n^T \tilde{\mathbf{w}}_k = \sum_{1 \le k \le n/2+1} X_{nk} \xrightarrow{d} \mathcal{N}\left(0, \|\boldsymbol{\alpha}\|_2^2\right)$$

By Cramér-Wold Theorem [36] $C_n^{-1/2} \mathbf{u}^{(n)} \xrightarrow{d} \mathcal{N}(0, I_r)$ $\square$.

*Proof of Theorem 3:* (a) Since $\pi$ is uniformly distributed, for any $1 \le k \le n$, $\pi(k)$ gets the values $\{1 \ldots n\}$ with equal probabilities. Therefore,

$$E\{w_{j\pi(k)}\} = n^{-1} \sum_{g=1}^{n} w_{jg} = \overline{w}_j,$$

$$E\{w_{j\pi(k)} w_{h\pi(k)}\} = n^{-1} \sum_{g=1}^{n} w_{jg} w_{hg} = n^{-1} \delta_{jh}.$$

Similarly, for any $1 \le k, l \le n, k \ne l$, the random pair $(\pi(k), \pi(l))$ gets the values $\{(g,p)\} \mid 1 \le g, p \le n, g \ne p\}$ with equal probabilities. Therefore, for $k \ne l$

$$E\{w_{j\pi(k)} w_{h\pi(l)}\} = \frac{1}{n(n-1)} \sum_{\substack{g,p=1 \\ p \ne g}}^{n} w_{jg} w_{hp} = \frac{1}{n(n-1)} \left[\left(\sum_{g=1}^{n} w_{jg}\right)\left(\sum_{p=1}^{n} w_{hp}\right) - \sum_{g=1}^{n} w_{jg} w_{hg}\right] = \frac{n^2 \overline{w}_j \overline{w}_h - \delta_{jh}}{n(n-1)}.$$

(34) follows from

$$E\{z_j\} = (n/m)^{1/2} \sum_{k=1}^{n} x_k E\{w_{j\pi(\kappa)}\} = n^{3/2} m^{-1/2} \overline{\mathbf{x}} \overline{\mathbf{w}}_j.$$

(35) and (36) are proved by computing $E\{z_j z_h\}$:

$$E\{z_j z_h\} = \frac{n}{m} \sum_{k,l=1}^{n} x_k x_l E\{w_{j\pi(k)} w_{h\pi(l)}\} = \frac{n}{m} \sum_{k=1}^{n} x_k^2 E\{w_{j\pi(k)} w_{h\pi(k)}\} + \frac{n}{m} \sum_{\substack{k,l=1 \\ k \ne l}}^{n} x_k x_l E\{w_{j\pi(k)} w_{h\pi(l)}\}$$

Using the results above and the identity $\sum_{\substack{k,l=1 \\ k \ne l}}^{n} x_k x_l = \left(\sum_{k=1}^{n} x_k\right)^2 - \sum_{k=1}^{n} x_k^2 = n^2 \overline{\mathbf{x}}^2 - \|\mathbf{x}\|_2^2$ we get

$$E\{z_j z_h\} = \frac{n}{m}\left[n^{-1} \delta_{jh} \|\mathbf{x}\|_2^2 + \frac{n^2 \overline{w}_j \overline{w}_i - \delta_{jh}}{n(n-1)}\left(n^2 \overline{\mathbf{x}}^2 - \|\mathbf{x}\|_2^2\right)\right] = \frac{n}{m(n-1)}\left[(\delta_{jh} - n \overline{w}_j \overline{w}_h)\|\mathbf{x}\|_2^2 + n(n^2 \overline{w}_j \overline{w}_h - \delta_{jh}) \overline{\mathbf{x}}^2\right] =$$

$$\frac{n}{m(n-1)}\left[(\delta_{jh} - n \overline{w}_j \overline{w}_h)(\|\mathbf{x}\|_2^2 - n \overline{\mathbf{x}}^2) + n^2(n-1) \overline{w}_j \overline{w}_h \overline{\mathbf{x}}^2\right] = \frac{n}{m(n-1)}(\delta_{jh} - n \overline{w}_j \overline{w}_h)(\|\mathbf{x}\|_2^2 - n \overline{\mathbf{x}}^2) + \frac{n^3}{m} \overline{w}_j \overline{w}_h \overline{\mathbf{x}}^2$$

(35) follows by substitution into (6) and (36) is obtained from (35) by setting $h = j$.

(b) Let

$$a_{kl}^{(j)} \triangleq \frac{(x_k - \bar{x})(w_{jl} - \bar{w}_j)}{2\|T\mathbf{w}_j^T\|_\infty \|T\mathbf{x}\|_\infty} + \frac{1}{2} = \frac{(T\mathbf{x})_k (T\mathbf{w}_j^T)_l}{2\|T\mathbf{w}_j^T\|_\infty \|T\mathbf{x}\|_\infty} + \frac{1}{2} \in [0,1]$$

$$S_j \triangleq \sum_{k=1}^n a_{k\pi(k)}^{(j)} = \frac{\sum_{k=1}^n w_{j\pi(k)} x_k - n\bar{w}_j \bar{x}}{\|T\mathbf{w}_j^T\|_\infty \|T\mathbf{x}\|_\infty} + \frac{n}{2}.$$

$E\{S_j\} = \sum_{k=1}^n E\{a_{k\pi(k)}^{(j)}\} = n/2$, hence $z_j - E\{z_j\} = (2\sigma_j/\tau_j)(S_j - E\{S_j\})$. Applying Chatterjee concentration inequality we get

$$\Pr\{|z_j - E\{z_j\}| \geq t\sigma_j\} = \Pr\{|S_j - E\{S_j\}| \geq t\tau_j/2\} \leq 2\exp[-(t\tau_j/2)^2/(2n + t\tau_j)] = 2\exp[-t^2 \eta_n(t, \tau_j)].$$

The inequality in (37) is because for any $\mathbf{v} \in \mathbb{R}^n$, $\|\mathbf{v}\|_2 \leq \sqrt{n}\|\mathbf{v}\|_\infty$.

(c) Let $C_n \triangleq \text{cov}(\mathbf{u}^{(n)})$. By (36) $(m/n)\sigma_{jn}^2 \leq n^{-1}\|\mathbf{x}^{(n)}\|_2^2 \leq x_{\max}^2$, hence by lemma 3 there is $\varepsilon > 0$ such that for sufficiently large $n$, $C_n$ is positive definite and for any $\boldsymbol{\alpha} \in \mathbb{R}^r$, $\boldsymbol{\alpha}^T C_n \boldsymbol{\alpha} \geq \varepsilon(n/m)\|\boldsymbol{\alpha}\|_2^2$. For any $\boldsymbol{\alpha} \in \mathbb{R}^r$, $\boldsymbol{\alpha} \neq \mathbf{0}$, and $1 \leq k \leq n$ define:

$$\tilde{\boldsymbol{\alpha}}_n \triangleq [\tilde{\alpha}_{n1}, \ldots, \tilde{\alpha}_{nr}]^T \triangleq C_n^{-1/2} \boldsymbol{\alpha}$$

$$\tilde{\mathbf{w}}_k^{(n)} \triangleq [w_{j_n(1),k}^{(n)}, \ldots, w_{j_n(r),k}^{(n)}]^T, \quad \bar{\tilde{\mathbf{w}}}^{(n)} \triangleq [\bar{w}_{j_n(1)}^{(n)}, \ldots, \bar{w}_{j_n(r)}^{(n)}]^T$$

$$a_n(k) \triangleq \sqrt{n/m}(x_k - \bar{x}^{(n)}), \quad b_n(k) \triangleq \tilde{\boldsymbol{\alpha}}_n^T (\tilde{\mathbf{w}}_k^{(n)} - \bar{\tilde{\mathbf{w}}}^{(n)}),$$

We show that the conditions of the combinatorial CLT hold for $\{a_n(k)\}, \{b_n(k)\}, 1 \leq k \leq n, n = 1, 2, \ldots$ Note that $\bar{a}_n = \bar{b}_n = 0$.

$$n^{-1}\sum_{k=1}^n (a_n(k) - \bar{a}_n)^2 \sum_{k=1}^n (b_n(k) - \bar{b}_n)^2 = m^{-1}\sum_{k=1}^n (x_k - \bar{x}^{(n)})^2 \sum_{k=1}^n (\tilde{\boldsymbol{\alpha}}_n^T \tilde{\mathbf{w}}_k^{(n)} - \tilde{\boldsymbol{\alpha}}_n^T \bar{\tilde{\mathbf{w}}}^{(n)})^2 =$$

$$m^{-1}\left(\|\mathbf{x}\|_2^2 - n(\bar{x}^{(n)})^2\right)\sum_{l,h=1}^r \tilde{\alpha}_{nl} \tilde{\alpha}_{nh} \left(\delta_{lh} - n\bar{w}_{j_n(l)}^{(n)} \bar{w}_{j_n(h)}^{(n)}\right) = (n-1)n^{-1}\tilde{\boldsymbol{\alpha}}_n^T C_n \tilde{\boldsymbol{\alpha}}_n \geq (n-1)m^{-1}\|\tilde{\boldsymbol{\alpha}}_n\|_2^2 \varepsilon \quad (42)$$

$$\max_{1 \leq k \leq n} (a_n(k) - \bar{a}_n)^2 = nm^{-1}\max_{1 \leq k \leq n}(x_k - \bar{x}^{(n)})^2 \leq 4nm^{-1}x_{\max}^2$$

$$\max_{1 \leq k \leq n}(b_n(k) - \bar{b}_n)^2 = \max_{1 \leq k \leq n}\left(\tilde{\boldsymbol{\alpha}}_n^T(\tilde{\mathbf{w}}_k^{(n)} - \bar{\tilde{\mathbf{w}}}^{(n)})\right)^2 \leq \max_{1 \leq k \leq n}\|\tilde{\boldsymbol{\alpha}}_n\|_2^2 \|\tilde{\mathbf{w}}_k^{(n)} - \bar{\tilde{\mathbf{w}}}^{(n)}\|_2^2 \leq \|\tilde{\boldsymbol{\alpha}}_n\|_2^2 \sum_{l=1}^r 4\|\mathbf{w}_{j_n(l)}^{(n)}\|_\infty^2$$

and, by (13)

$$n \frac{\max_{1 \leq k \leq n}(a_n(k) - \bar{a}_n)^2}{\sum_{k=1}^n (a_n(k) - \bar{a}_n)^2} \frac{\max_{1 \leq k \leq n}(b_n(k) - \bar{b}_n)^2}{\sum_{k=1}^n (b_n(k) - \bar{b}_n)^2} \leq 16n(n-1)^{-1}x_{\max}^2 \varepsilon^{-1} \sum_{l=1}^r \|\mathbf{w}_{j_n(l)}^{(n)}\|_\infty^2 \xrightarrow[n \to \infty]{} 0$$

Therefore, condition (41) holds, and since the right hand side of (42) is positive the sequences $\{a_n(k)\}$, $\{b_n(k)\}$, $1 \leq k \leq n$ are non-constant for sufficiently large $n$. Hence $\sum_{k=1}^n a_n(k) b_n(\pi(k))$ is asymptotically normal. Let $\tilde{\mathbf{u}}^{(n)} \triangleq C_n^{-1/2}(\mathbf{u}^{(n)} - E\{\mathbf{u}^{(n)}\})$.

$E\{\tilde{\mathbf{u}}^{(n)}\} = \mathbf{0}$, $E\{\tilde{\mathbf{u}}^{(n)}\tilde{\mathbf{u}}^{(n)T}\} = I_r$ and

$$\boldsymbol{\alpha}^T\tilde{\mathbf{u}}^{(n)} = \sqrt{n/m}\tilde{\boldsymbol{\alpha}}_n^T\left[\sum_{k=1}^n x_k \tilde{\mathbf{w}}_{\pi(k)}^{(n)} - n\overline{\mathbf{x}}^{(n)}\overline{\tilde{\mathbf{w}}}^{(n)}\right] = \sqrt{n/m}\tilde{\boldsymbol{\alpha}}_n^T \sum_{k=1}^n \left(x_k - \overline{\mathbf{x}}^{(n)}\right)\left(\tilde{\mathbf{w}}_{\pi(k)}^{(n)} - \overline{\tilde{\mathbf{w}}}^{(n)}\right) = \sum_{k=1}^n a_n(k) b_n(\pi(k))$$

By Cramér-Wold Theorem [36]

$$C_n^{-1/2}(\mathbf{u}^{(n)} - E\{\mathbf{u}^{(n)}\}) = \tilde{\mathbf{u}}^{(n)} \xrightarrow{d} \mathcal{N}(0, I_r) \quad \square.$$

## ACKNOWLEDGMENT

We are grateful to Hong Jiang, Jin Cao, Tin Ho and Alexei Ashikhmin of Bell-Laboratories for helpful discussions and suggestions.

## REFERENCES

[1] M. A. Davenport, F Duarte, Marco, Y. C. Eldar, and G. Kutyniok, "Introduction to Compressed Sensing," in *Compressed Sensing*, Y.C. Eldar and G. Kutyniok, Eds., Cambridge, UK: Cambridge University Press, 2012.

[2] R. Calderbank, S. Howard, and S. Jafarpour, "Construction of a Large Class of Deterministic Sensing Matrices That Satisfy a Statistical Isometry Property," *IEEE J. Sel. Top. Signal Process.*, vol. 4, no. 2, pp. 358–374, Apr. 2010.

[3] E. J. Candès and T. Tao, "Near-Optimal Signal Recovery From Random Projections: Universal Encoding Strategies?," *IEEE Transactions on Information Theory*, vol. 52, no. 12, pp. 5406–5425, Dec. 2006.

[4] E. J. Candès and T. Tao, "Decoding by Linear Programming," *IEEE Trans. Inf. Theory*, vol. 51, no. 12, pp. 4203–4215, Dec. 2005.

[5] E. Candès and J. Romberg, "Sparsity and incoherence in compressive sampling," *Inverse Probl.*, vol. 23, no. 3, pp. 969–985, Jun. 2007.

[6] T. T. Do, T. D.Tran, and L. Gan, "Fast compressive sampling with structurally random matrices," in *ICASSP 2008 IEEE International Conference Acoustic Speech & Signal Processing*, 2008, pp. 3369–3372.

[7] T. T. Do, L. Gan, N. H. Nguyen, and T. D. Tran, "Fast and Efficient Compressive Sensing Using Structurally Random Matrices," *IEEE Transactions on Signal Processing*, vol. 60, no. 1, pp. 139–154, Jan. 2012.

[8] J. Romberg, "Compressive Sensing by Random Convolution," *SIAM J. Imaging Sci.*, vol. 2, no. 4, pp. 1098–1128, Jan. 2009.

[9] F. Krahmer and R. Ward, "New and Improved Johnson–Lindenstrauss Embeddings via the Restricted Isometry Property," *SIAM J. Math. Anal.*, vol. 43, no. 3, pp. 1269–1281, Jan. 2011.

[10] C. Li, H. Jiang, P. Wilford, and Y. Zhang, "Video coding using compressive sensing for wireless communications," *2011 IEEE Wireless Communications and Networking Conference*, pp. 2077–2082, Mar. 2011.

[11] C. Li, H. Jiang, P. Wilford, Y. Zhang, and M. Scheutzow, "A new compressive video sensing framework for mobile broadcast," *IEEE Transactions on Broadcasting*, vol. 59, no. 1, pp. 197–205, Mar. 2013.

[12] S. Pudlewski and T. Melodia, "On the performance of compressive video streaming for wireless multimedia sensor networks," in *2010 IEEE International Conference on Communications*, 2010, pp. 1-5.

[13] H. Jiang, C. Li, R. Haimi-Cohen, P. A. Wilford, and Y. Zhang, "Scalable video coding using compressive sensing," *Bell Labs Technical Journal*, vol. 16, no. 4, pp. 149–169, Mar. 2012.

[14] D. Venkatraman and A. Makur, "A compressive sensing approach to object-based surveillance video coding," in *2009 IEEE International Conference on Acoustics, Speech and Signal Processing*, 2009, pp. 3513–3516.

[15] A. Mahalanobis and R. Muise, "Object specific image reconstruction using a compressive sensing architecture for application in surveillance systems," *IEEE Transactions on Aerospace and Electronic Systems*, vol. 45, no. 3, pp. 1167–1180, Jul. 2009.

[16] H. Jiang, S. Zhao, Z. Shen, W. Deng, P. Wilford, and R. Haimi-Cohen, "Video Analysis using Compressive Sensing with Low Latency," *Bell Labs Tech. J.*, vol. 18, no. 4, 2014.

[17] J. Makhoul, "Linear prediction: a tutorial," Proc. IEEE, vol. 63, pp. 561-580, Apr. 1975.

[18] A. M. Kondoz, *Digital Speech*, Sec. 3.3, 4.2. Chichester, 1994.

[19] W. Dai, H. V. Pham, and O. Milenkovic, "Distortion-rate functions for quantized compressive sensing," in *2009 IEEE Information Theory Workshop on Networking and Information Theory*, 2009, pp. 171–175.

[20] W. Dai, R. V. Pham, and O. Milenkovic, "A comparative study of quantized compressive sensing schemes," in *2009 IEEE International Symposium on Information Theory*, 2009, pp. 11–15.

[21] Y. Baig, E. M.-K. Lai, and J. P. Lewis, "Quantization effects on compressed sensing video," in *2010 17th International Conference on Telecommunications*, 2010, pp. 935–940.

[22] J. Z. Sun and V. K. Goyal, "Optimal quantization of random measurements in compressed sensing," *2009 IEEE International Symposium on Information Theory*, vol. 3, no. 1, pp. 6–10, Jun. 2009.

[23] J. N. Laska, P. T. Boufounos, M. a. Davenport, and R. G. Baraniuk, "Democracy in action: Quantization, saturation, and compressive sensing," *Appl. Comput. Harmon. Anal.*, vol. 31, no. 3, pp. 429–443, Nov. 2011.

[24] J. N. Laska and R. G. Baraniuk, "Regime change: Bit-depth versus measurement-rate in compressive sensing," *IEEE Transactions on Signal Processing*, vol. 60, no. 7, pp. 3496–3505, Jul. 2012.

[25] J. N. Laska and R. G. Baraniuk, "Trust, But Verify: Fast and Accurate Signal Recovery From 1-Bit Compressive Measurements," *IEEE Trans. Signal Process.*, vol. 59, no. 11, pp. 5289–5301, Nov. 2011.

[26] A. Zymnis, S. Boyd, and E. Candès, "Compressed sensing with quantized measurements," *IEEE Signal Processing Letters*, vol. 17, no. 2, pp. 149–152, 2010.

[27] L. Jacques, D. K. Hammond, and J. M. Fadili, "Dequantizing compressed sensing : When constraints combine," *IEEE Transaction on Information Theory*, vol. 57, no. 1, pp. 559–571, 2011.

[28] M. F. Duarte and R. G. Baraniuk, "Kronecker compressive sensing.," *IEEE Trans. Image Process.*, vol. 21, no. 2, pp. 494–504, Feb. 2012.

[29] J. Rissanen and G. G. Langdon, "Arithmetic coding," *IBM J. Res. Dev.*, vol. 23, no. 2, pp. 149–162, 1979

[30] A. Gersho and R. M. Gray, *Vector Quantization and Signal Compression* Kluwer Academic Publishers, 1992, sec. 9.9, pp. 295-302.

[31] J. Ziv, "On universal quantization," *IEEE Transactions on Information Theory*, vol. 31, no. 3, pp. 344–347, May 1985.

[32] M. F. Duarte, M. A. Davenport, D. Takhar, J. N. Laska, T. Sun, K. F. Kelly, and R. G. Baraniuk, "Single-Pixel Imaging via Compressive Sampling," *IEEE Signal Process. Mag.*, vol. 25, no. 2, pp. 83–91, 2008.

[33] H.C. Thode, Testing for Normality New York: Marcel Dekker Inc., 2002, pp. 21-23.

[34] W. Hoeffding, "Probability Inequalities for Sums of Bounded Random Variables," *J. Am. Stat. Assoc.*, vol. 58, no. 301, pp. 13–30.

[35] G. Bennett, "Probability inequalities for the sum of independent random variables," J. American Stat. Assoc., vol. 57(297), pp. 33-45, 1962

[36] P. Billingsley, *Probability and Measure*, 3rd ed. New York:Wiley, 1995.

[37] S. Chatterjee, "Stein's method for concentration inequalities," Probability Theory and Related Fields, vol. 138, no. 1–2, pp. 305–321, Apr. 2007.

[38] W. Hoeffding, "A Combinatorial Central Limit Theorem," The Annals of Mathematical Statistics, vol. 22, no. 4, pp. 558–566, 1951.

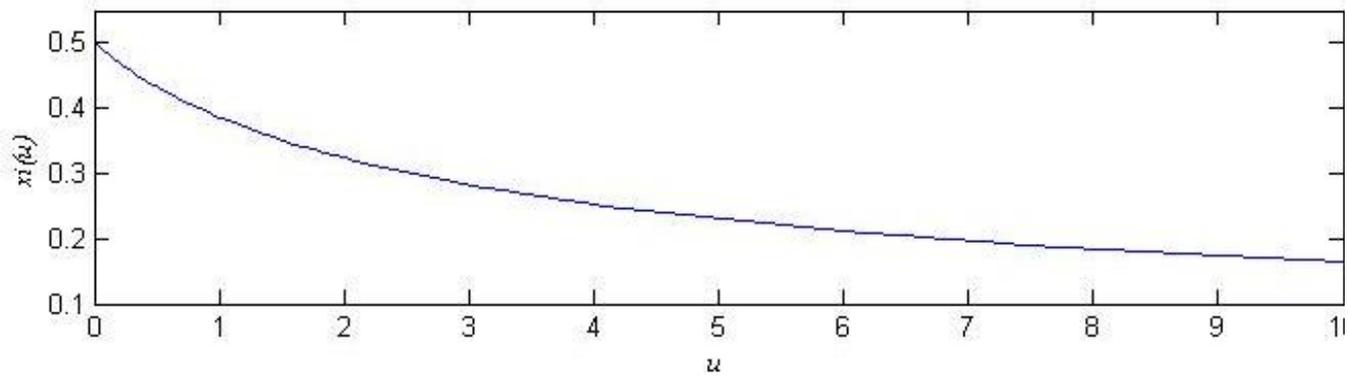

Fig. 1. The function $\xi(u)$ of (31)

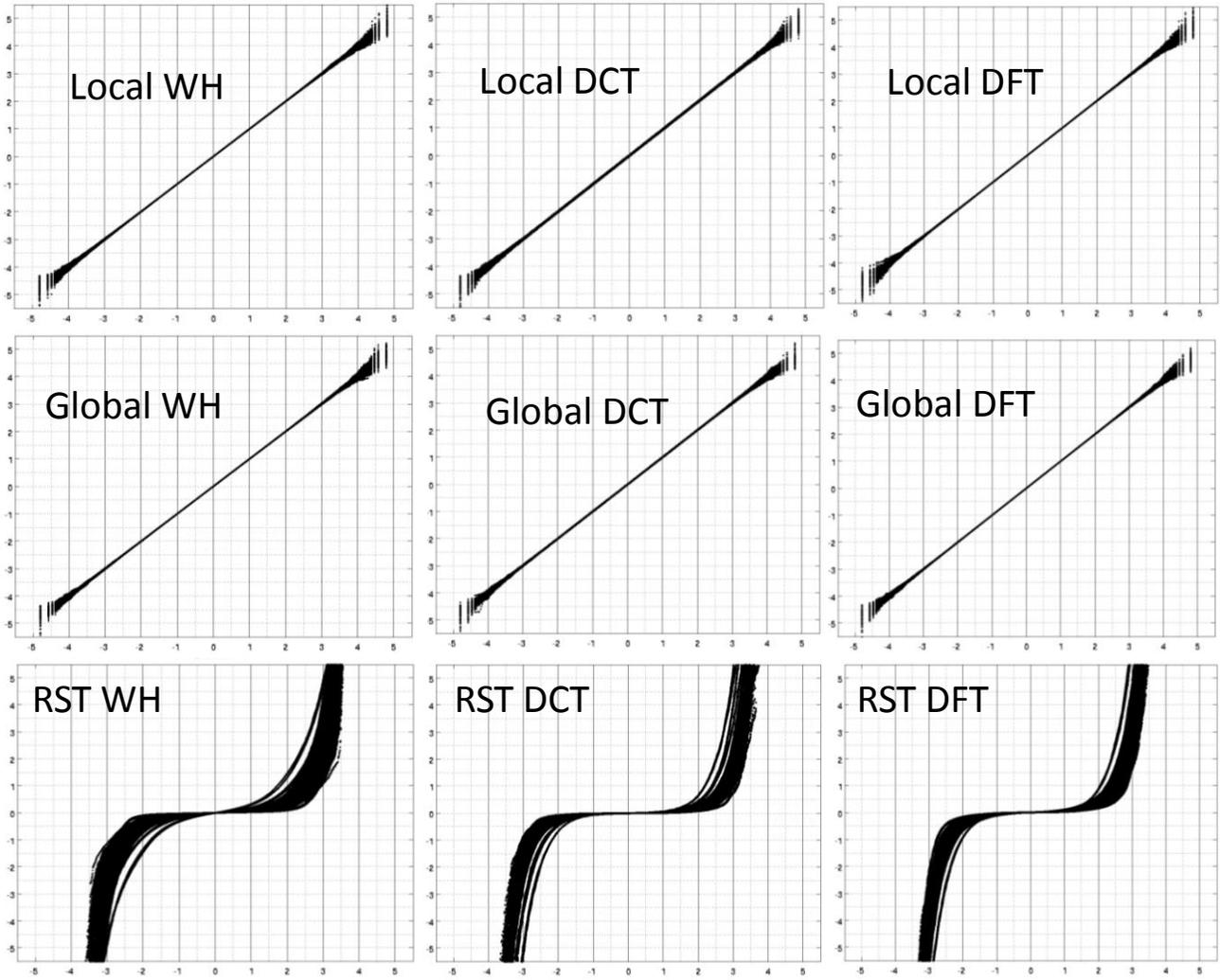

Fig. 2. Q-Q plots, in the range [-5,5], of measurements generated by different sensing matrices. First row: LR-SRMs; second row: GR-SRMs; third row: RSTs (no randomization). Each graph shows 160 overlaid Q-Q plots, each plot comprising 76031 measurements.

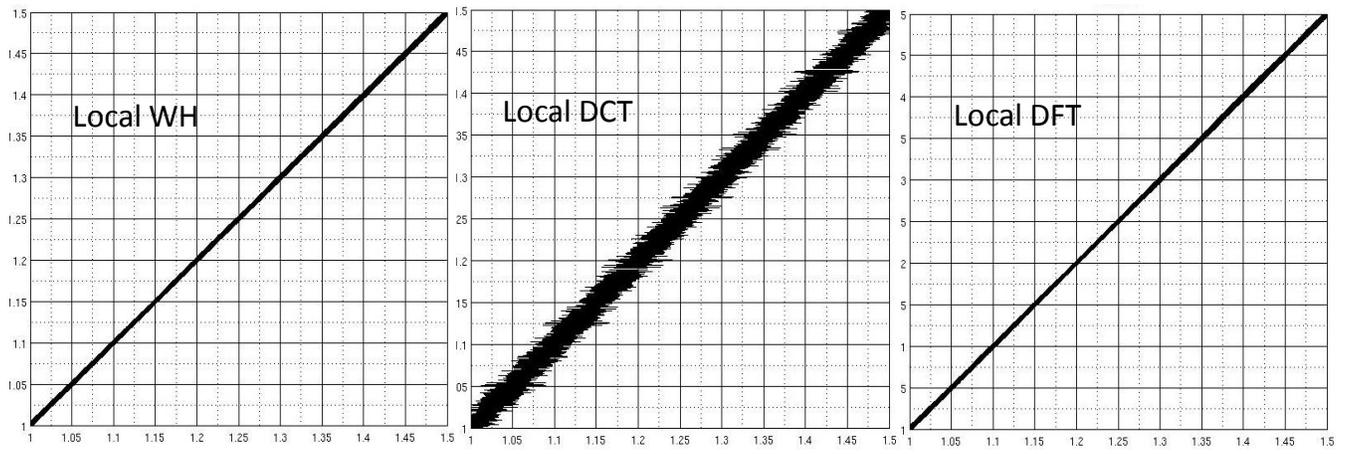

Fig. 3. Enlargements the range [1,1.5] in the Q-Q plots of measurements generated with SRMs with local randomization.